\documentclass[
superscriptaddress,
onecolumn,
prd,tightenlines,
nofootinbib, eqsecnum,
amsfonts,amsmath,amssymb]{revtex4}
\usepackage{bm}
\usepackage{dsfont}
\usepackage{graphicx} 
\usepackage{hyperref}
\usepackage{mathrsfs}
\usepackage{multirow}

\newcommand{\ud}{\mathrm{d}}

\usepackage{ulem}
\usepackage[usenames]{color}

\newcommand{\red}{\textcolor{red}}

\usepackage{textcomp}
\usepackage{url}
\usepackage{soul}
\usepackage{natbib}
\usepackage{siunitx}
\newcommand{\Eotvos}{E\"otv\"os}
\usepackage{wrapfig}

\allowdisplaybreaks
\begin{document}

%\title{Exploring the Foundations of the Universe with\\ Space Tests of the Equivalence Principle}
%
%\date{\today}
%
%
%
%\maketitle

%%%%%%%%%%%%%%%%%%%%%%%%%%%%%%%%%%%%%

%\begin{enumerate}
%\item Cover page with title and contact information - 1 page - SYRTE
%\item Back cover page with list of proposers and affiliations - 1 page - SYRTE
%\item Executive summary - 1 page max. - SYRTE
%\item Testing the Universality of Free Fall (UFF) - scientific motivation - 6 pages max. - SYRTE
%\item Secondary science objectives - 1 page max. - SYRTE
%\item Technological Impact - 0.5 pages max. - ZARM
%\item An Atom Interferometric space test of UFF - 6 pages max. - LUH
%\item An advanced MICROSCOPE mission - 6 pages max. - ONERA
%\item Bibliography - 6 pages max. - All
%\item Conclusion
%\end{enumerate}
%
%\newpage

\begin{center}
White paper for the ESA Voyage 2050 long term plan \\
August 5, 2019

\vspace{7cm}

{\huge \bf Exploring the Foundations of the \\
\vspace{0.5cm}
Physical Universe with Space Tests of the \\
\vspace{0.5cm}
Equivalence Principle} 
%\vspace{0.7cm}
%Space Tests of the Equivalence Principle}

\vspace{4cm}

Peter Wolf \\
SYRTE, Observatoire de Paris, Universit\'e PSL, CNRS, Sorbonne Universit\'e, LNE \\
61 avenue de l'Observatoire \\
75014 Paris, France \\
e-mail: peter.wolf@obspm.fr

\end{center}

\newpage

\noindent{\bf Proposing team:}
%\footnote{When several scientists from the same institute are participating only the lead contact is given, with supporting colleagues in brackets.}} \\

\vspace{1cm}

\noindent Luc Blanchet, GRECO, Institut d’Astrophysique de Paris, CNRS, Sorbonne Universit\'e, FRANCE \\

\noindent Kai Bongs, School of Physics and Astronomy, University of Birmingham, UNITED KINGDOM \\

\noindent Philippe Bouyer, B. Battelier, A. Bertoldi, LP2N, IOGS, CNRS, Universit\'e de Bordeaux, Institut d’Optique, FRANCE \\

\noindent Claus Braxmaier, L. W\"orner, DLR Institute of Space Systems, System Enabling Technologies, and Universit\"at Bremen, ZARM, Center of applied space technology and microgravity, GERMANY \\

\noindent Davide Calonico, Istituto Nazionale di Ricerca Metrologica (INRIM), ITALY \\

\noindent Pierre Fayet, LPTENS, Ecole normale sup\'erieure, PSL, CNRS, Sorbonne Universit\'e, Universit\'e de Paris, FRANCE \\

\noindent Aur\'elien Hees, C. Le Poncin-Lafitte, P. Tuckey, SYRTE, Observatoire de Paris, Universit\'e PSL, CNRS, Sorbonne Universit\'e, LNE, FRANCE \\ 

\noindent Philippe Jetzer, Physik Institut, Universit\"at Z\"urich, SWITZERLAND \\ 

\noindent Claus L\"ammerzahl, Universit\"at Bremen, ZARM, Center of applied space technology and microgravity, GERMANY \\

\noindent Steve Lecomte, Centre Suisse d'Electronique et de Microtechnique (CSEM), SWITZERLAND \\

\noindent Gilles M\'etris, Universit\'e C\^ote d’Azur, Observatoire de la C\^ote d’Azur, CNRS, IRD, G\'eoazur, FRANCE \\

\noindent Ernst Rasel, N. Gaaloul, S. Loriani, C. Schubert, Institut f\"ur Quantenoptik and Centre for Quantum Engineering and Space-Time Research (QUEST), Leibniz Universit\"at Hannover, GERMANY \\

\noindent Serge Reynaud, C. Guerlin, C. Salomon, Laboratoire Kastler Brossel, CNRS, Sorbonne Universit\'e, ENS-PSL Universit\'e, Coll\`ege de France, FRANCE\\

\noindent Manuel Rodrigues, J. Berg\'e, D\'partement de Mesures Physiques, ONERA, FRANCE\\

\noindent Markus Rothacher, Institute of Geodesy and Photogrammetry, ETH Z\"urich, SWITZERLAND \\

\noindent Wolfgang P. Schleich, A. Roura, Institut f\"ur Quantenphysik, Universit\"at Ulm, GERMANY \\

\noindent Stephan Schiller, Institut f\"ur Experimentalphysik, Heinrich-Heine-Universit\"at D\"usseldorf, GERMANY \\

\noindent Carlos Sopuerta, M. Nofrarias, Institute of Space Sciences, (ICE, CSIC) and Institute of Space Studies of Catalonia (IEEC), SPAIN \\

\noindent Fiodor Sorrentino, Istituto Nazionale di Fisica Nucleare, INFN - sez. di Genova, ITALY \\

\noindent Tim J. Sumner, High Energy Physics Group, Imperial College London, Blackett Laboratory, UNITED KINGDOM \\ 

\noindent Guglielmo M. Tino, Dipartimento di Fisica e Astronomia and LENS Laboratory, Universit\`a degli Studi di Firenze, INFN - sez. di Firenze, ITALY \\

\noindent Wolf von Klitzing, Inst. of Electronic Structure and Laser, Foundation for Research and Technology-Hellas, GREECE \\

\noindent Martin Zelan,  Measurement Science and Technology, RISE Research Institutes of Sweden, SWEDEN

\newpage
\section{Executive Summary}

Einstein's theory of general relativity (GR) is a cornerstone of our current description of the physical world. It is used to understand the flow of time in presence of gravity, the motion of bodies from satellites to galaxy clusters, the propagation of electromagnetic waves in the presence of massive bodies, the generation and propagation of gravitational waves, the evolution of stars, and of the Universe as a whole.

But there is a strong asymmetry between GR and the other interactions of the standard model of particle physics (SM): the electromagnetic, weak, and strong interactions. Whilst the latter couple to some specific property or charge, gravitation is universally coupled, meaning that it couples in the same way to any mass/energy, which allows a geometric description of gravitation as the effect of the curvature of space-time. The phenomenological manifestation of this universal coupling is known as the Einstein Equivalence Principle (EEP), and is central to modern physics at all scales.

The EEP is not a fundamental symmetry, like {\it e.g.} gauge invariance in the SM, but rather an experimental fact. Einstein himself initially called it the \textit{hypothesis of equivalence} before elevating it to a \textit{principle} once it became clear how central it was in the
generalization of special relativity to include gravitation. And indeed, from a SM perspective it is rather surprising that the EEP should be satisfied at all, let alone at the stringent uncertainties of present-day tests. Furthermore, the difficulties in quantizing GR and in unifying it with the SM gives further indications that the EEP must be violated at some level. For example, most attempts at unification theories involve additional fields, that have no good reason to couple universally to the SM and thus would violate the EEP. Similarly, the unknown nature of dark energy and dark matter postulated by modern cosmology and astronomy, is often ``explained'' by invoking additional fields that permeate space-time. Again such fields would in general couple non-universally to the SM and thus violate the EEP.

\begin{wrapfigure}{r}{0.5\textwidth}
  \begin{center}
    \includegraphics[width=0.48\textwidth]{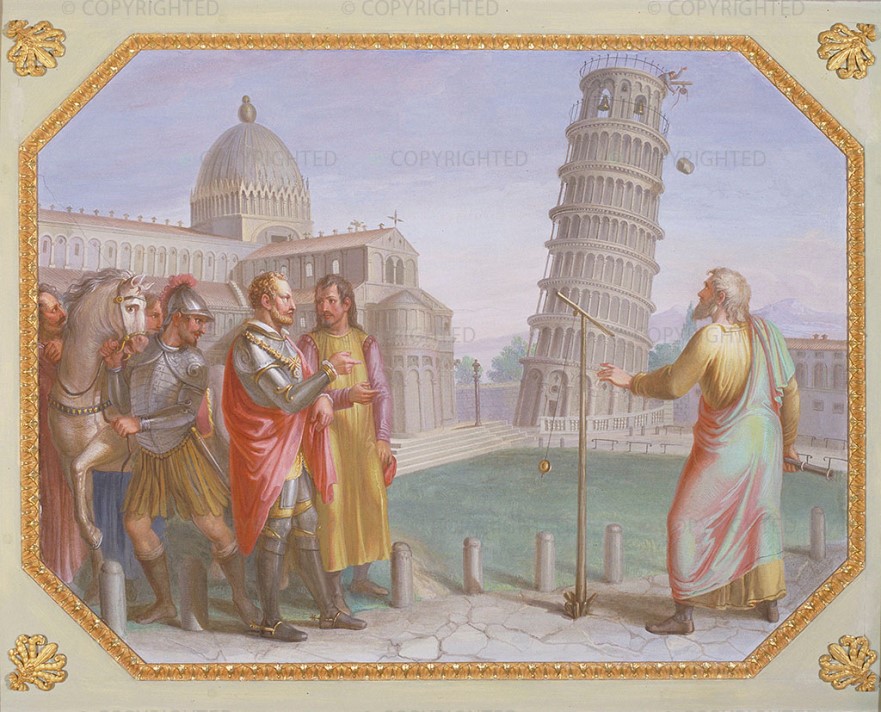}
  \end{center}
  \caption{{\it In the presence of the Grand Duke, Galileo Galilei performs the experiment of falling bodies from the Tower of Pisa}. Fresco by Luigi Catani, 1816 (Firenze, Palazzo Pitti, Quartiere Borbonico o Nuovo Palatino, sala 15). Whilst the historical veracity of this particular experiment is debatable, the fact that Galileo was one of the first scientists to carry out UFF tests is well established.}
\end{wrapfigure}

These considerations, detailed in the main text, make experimental tests of the EEP one of the most promising roads to discovering new physics beyond the SM and GR. By doing so one may shed new light on much of our present day understanding of the universe, and in particular its main constituents, cold dark matter and dark energy, both of which we know nothing about apart from their gravitational manifestations. Additionally, diversifying the tests by using new forms of test-masses {\it e.g.} atoms in quantum superpositions, may give access to the interplay between the SM and GR at the most fundamental level.

Exploring the extent to which the EEP is satisfied is then the main subject of this white paper. Finding a violation of the EEP would not only revolutionise physics as a whole, but certainly also shed new light on astrophysics and cosmology, in particular concerning its dark components.

The history of experimental tests of the EEP dates back at least as far as the 16th century and Galileo Galilei. Since then, tremendous efforts have been carried out to push laboratory tests to uncertainties as low as parts in $10^{-13}$ when testing the universality of free fall (UFF)\footnote{Tests of UFF are generally quantified by the \Eotvos-ratio defined as $\eta = 2(a_1 - a_2)/(a_1+a_2)$ where $a_{1,2}$ are the gravitational accelerations of two test masses of different composition.}, the best known aspect of the EEP. However, ground tests are ultimately limited by the Earth's gravitational environment, and future discoveries will come from space experiments, like the recent MICROSCOPE experiment, which between 2016 and 2018 tested the UFF in space. First partial result excluded a violation of the EEP at the $10^{-14}$ level \cite{Touboul2017} whilst final results (expected later this year) will search for a violation down to the low $10^{-15}$ region.

Over the last years, many proposals for space-tests of UFF have been put forward, e.g. STEP \cite{Sumner2007}, GG \cite{Nobili2012}, POEM \cite{Reasenberg2014}, GAUGE \cite{Amelino2009}, STE-QUEST \cite{Aguilera2014,Altschul2015}, and the future will certainly be built on these and the heritage of MICROSCOPE.

In the post-MICROSCOPE era, subject of this white paper, the aim will be to either confirm the discovery of a UFF violation by MICROSCOPE at the low $10^{-15}$ level, and/or to explore further in terms of sensitivity and diversity of test masses. The uncertainties aimed for will be $\leq 10^{-17}$, a leap in sensitivity by more than two orders of magnitude. 

Two mission concepts that can achieve that goal are presented, one based on cold-atom technology following the STE-QUEST proposal, the other based on an evolution of the MICROSCOPE technology. Both concepts are expected to fit into the M-class envelope, or possibly smaller mission profiles. The mission scenarios are at this stage only tentative, and feature low Earth orbits with drag-free technology as convincingly demonstrated by MICROSCOPE and LISA-Pathfinder \cite{Armano2018,Armano2019}.

The main technological challenges are discussed in the respective sections. Here we only point out the strong technology development activities that have been ongoing over the last years, in particular in the context of cold atom interferometry in microgravity through the QUANTUS, MAIUS projects \cite{Muntinga2013,Becker2018} in Germany and the ICE project in France \cite{Geiger2011}. Additionally, recent theoretical and experimental results have allowed to strongly reduce some of the main systematic effects in such experiments \cite{Roura2017,Overstreet2017,Damico2017}. 

In summary, future space tests of the UFF, and more generally the EEP, are one of our best hopes for a major discovery that will revolutionise not only fundamental physics, but also our understanding of the universe at all scales and in particular the present day enigmas of dark energy and cold dark matter. Europe has a clear lead in this field, through recent missions like MICROSCOPE \cite{Touboul2017} and LISA-Pathfinfder \cite{Armano2018} that are unique in the history of space-science, and through upcoming missions like ACES \cite{Cacciapuoti2009}. It is now time to build on that heritage and pave the way towards the future, which may well lead to ground breaking new discoveries for mankind.

\section{Scientific Motivation}

Our best knowledge of the physical Universe, at the deepest
fundamental level, is based on two theories: Quantum Mechanics (or,
more precisely, Quantum Field Theory) and the classical theory of
General Relativity. 

Quantum Field Theory has been extremely successful
in providing an understanding of the observed phenomena of atomic,
particle, and high energy physics and has allowed a coherent
description of three of the four fundamental interactions that are
known to us: electromagnetic, weak and strong interactions (the fourth
one being gravitation). It has led to the Standard Model (SM) of particle
physics that has been highly successful in interpreting most observed
particle phenomena, and has been strongly confirmed with the discovery at the LHC of the Higgs (or, more precisely,
Brout-Englert-Higgs) boson, which could in fact be viewed as the
discovery of a fifth fundamental interaction. Although open questions
remain within the SM, it is clearly
the most compelling model for fundamental interactions at the
microscopic level that we have at present.

On the other hand, Einstein's theory of General Relativity (GR) is a
cornerstone of our current description of the physical world at
macroscopic scales. It is used to understand the flow of time in the
presence of gravity, the motion of bodies from satellites to galaxy
clusters, the propagation of electromagnetic waves in the vicinity of
massive bodies, the generation and propagation of gravitational waves, the evolution of stars, and the dynamics of the
Universe as a whole. It has most recently been confirmed by the direct detection in LIGO and Virgo of gravitational waves from mergers of black holes or neutron stars. GR brilliantly accounts for all observed
phenomena related to gravitation, in particular all observations in
the Earth's environment, the Solar system, and on galactic and cosmological scales.

The assumed validity of GR at cosmological scales, and the fact that
non-gravitational interactions are described by the SM, together with a hypothesis of homogeneity and
isotropy of cosmological solutions of these theories, have led to the
``concordance model'' of cosmology, referred to as the $\Lambda$-CDM model, which is in agreement with all present-day
observations at large scales, notably the most recent observations of
the anisotropies of the cosmic microwave background by the Planck
satellite~\cite{Ade2013}. However, important difficulties remain, in
particular the necessary introduction of dark energy, described by a
cosmological constant $\Lambda$ whose tiny measured value remains
unexplained so far, and of cold dark matter (CDM), made of some
unknown, yet to be discovered, stable particle.

There is a potential conflict on the problem of dark matter between
the concordance model of cosmology and the SM. On the one hand, there is strong evidence~\cite{Ade2013}
that 26.8 \% of the mass-energy of the Universe is made of
non-baryonic dark matter particles, which should certainly be
accounted for by some extension of the SM. On the
other hand, there is no clear indication of new physics beyond the SM which has been found at the LHC or elsewhere.

Although very successful so far, GR as well as numerous other
alternative or more general theories of gravitation are classical
theories. As such, we expect that they are fundamentally incomplete, because they do
not include quantum effects. A theory solving this problem would
represent a crucial step towards the unification of all fundamental
forces of Nature. Most physicists believe that GR and the SM are only low-energy approximations of a more
fundamental theory that remains to be discovered. Several concepts
have been proposed and are currently under investigation
(\textit{e.g.}, string theory, loop quantum gravity, extra spatial
dimensions) to bridge this gap and most of them lead to tiny
violations of the basic principles of GR.

One of the most desirable attributes of that fundamental theory is the
unification of the fundamental interactions of Nature, \textit{i.e.} a
unified description of gravity and the three other fundamental
interactions. Several attempts at formulating such a theory have been made,
but none of them is widely accepted and considered
successful. Furthermore, they make very few precise quantitative
predictions that could be verified experimentally.
% One of them is the
%Hawking radiation of black holes, which is however far from being
%testable experimentally for stellar-size black holes we observe in
%astrophysics.

A central point in this field is, that most unification theories have
in common a violation at some (\textit{a priori} unknown) level of one
of the basic postulates of GR, which can be tested experimentally: the
Einstein Equivalence Principle (EEP). Let us emphasize that the EEP is
not a fundamental symmetry of physics, contrary to \textit{e.g.} the
principle of local gauge invariance in particle physics. Indeed, any new field introduced by an extension to the SM has no good reason to be universally coupled to the SM fields, thus leading to an apparent violation of the EEP. An important challenge is therefore to test the EEP with the best possible accuracy. This is then the main motivation of many experiments in
fundamental physics, both on Earth and in space.

Precision measurements are at the heart of the scientific method that,
since Galileo's time, is being used for unveiling Nature and
understanding its fundamental laws. The assumptions and predictions of
GR can be challenged by precision experiments on scales ranging from
micrometers in the laboratory to the Solar System, in the latter case
using spacecrafts or the orbiting Earth, Moon and planets. Advances in precision and diversity of the measurements leads to new discoveries and improved and diversified tests of the EEP. The implementation of tests with significantly improved sensitivity
obviously requires the use of state-of-the-art technology, and in case
of satellite-based experiments the challenge is to make such
technology compatible with use in space, \textit{i.e.} extremely
robust, reliable, and automatized.

\subsection{The Einstein Equivalence Principle}
\label{sec:EEP}

The foundations of gravitational theories and the equivalence
principle have been clarified by many authors, including
Schiff~\cite{Schiff1960}, Dicke~\cite{Dicke1964}, Thorne, Lee \&
Lightman~\cite{Thorne1973}, and others. Following the book of
Will~\cite{Will1993} the EEP is generally divided into three
sub-principles: the Weak Equivalence Principle (WEP) also known as the
Universality of Free Fall (UFF), Local Lorentz Invariance (LLI), and
Local Position Invariance (LPI). The EEP is satisfied if and only if
all three sub-principles are satisfied. Below we describe these three
sub-principles:
\begin{enumerate}
\item WEP (or UFF) states that if any uncharged test body\footnote{By
  test body is meant an electrically neutral body whose size is small
  enough that the coupling to inhomogeneities in the gravitational
  field can be neglected.} is placed at an initial event in space-time
  and given an initial velocity there, then its subsequent trajectory
  will be independent of its internal structure and composition. The
  most common test of WEP consists in measuring the relative
  acceleration of two test bodies of different internal structure and
  composition freely falling in the same gravitational field. If WEP
  is satisfied, their differential acceleration is zero;
\item LLI states that the outcome of any local non-gravitational test
  experiment is independent of the velocity and orientation of the
  (freely falling) apparatus. Tests of LLI usually involve a local
  experiment (\textit{e.g.} the comparison of the frequency of two
  different types of clocks) whose velocity and/or orientation is
  varied in space-time. LLI is verified if the result of the
  experiment is unaltered by that variation;
\item LPI states that the outcome of any local non-gravitational test
  experiment is independent of where and when in the Universe it is
  performed. Tests of LPI usually involve a local experiment
  (\textit{e.g.} the measurement of a fundamental constant, or the
  comparison of two clocks based on different physical processes) at
  different locations and/or times. In particular, varying the local
  gravitational potential allows for searches of some anomalous
  coupling between gravity and the fields involved in the local
  experiment. A particular version of such tests, known as test of the
  gravitational red-shift, uses the same type of clock, but at two
  different locations (different local gravitational potentials) and
  compares them \textit{via} an electromagnetic signal. Then it can be
  shown (see Sec.~2.4c in Ref.~\cite{Will1993}) that the measured
  relative frequency difference is equal to $\Delta U/c^2$ (where
  $\Delta U$ is the difference in gravitational potential) if and only
  if LPI is satisfied.
\end{enumerate}
Since the three sub-principles described above are very different in
their empirical consequences, it is tempting to regard them as
independent. However, it was realized quite early that any
self-consistent gravitational theory is very likely to contain
connections between the three sub-principles. This has become known as
Schiff's conjecture~\cite{Schiff1960}, formulated around 1960. Loosely
stated, Schiff's conjecture implies that if one of the three
sub-principles is violated, then so are the other two. Schiff's conjecture has given rise to much debate, in particular
concerning its empirical consequences and the relative merit of tests
of the different sub-principles. Whilst it is true that any theory
respecting energy conservation (\textit{e.g.} based on an invariant
action principle) must satisfy Schiff's conjecture, the actual
quantitative relationship between violation of the sub-principles is
model dependent and varies as a function of the mechanism used for the
violation. As a consequence, it is not known \textit{a
  priori} which test (WEP/UFF, LLI, or LPI) is more likely to first
detect a violation and the most reasonable approach is to perform the
tests of the three sub-principles at the best possible precision.

For completeness, and to avoid possible confusion, we will say a few
words about the Strong Equivalence Principle (SEP), although it is not
directly related to this white paper. The SEP is a
generalization of EEP to include ``test'' bodies with non-negligible
self-gravitation, together with experiments involving gravitational
forces (\textit{e.g.} Cavendish-type experiments) or the propagation of gravitational waves. Obviously, SEP
includes EEP as a special case in which gravitational forces can be
ignored. Typical tests of SEP involve moons, planets, stars or local
gravitational experiments, the best known example being lunar laser
ranging that tests the universality of free fall, with the two test
bodies being the Moon and the Earth falling in the field of the
Sun. Clearly the two test bodies have non-negligible self-gravitation
and thus provide a test of SEP. The empirical consequences of SEP and
EEP are quite different; in general a violation of SEP does not
necessarily imply a violation of EEP. Similarly the theoretical
consequences are very different: a violation of EEP excludes not only
GR as a possible theory of gravitation, but also all other metric
theories (\textit{e.g.} all PPN theories, Brans-Dicke theory,
\textit{etc.}). A violation of SEP on the other hand excludes GR, but
allows for a host of other metric theories (\textit{e.g.} PPN theories
that satisfy a particular combination of PPN parameters). In that
sense, SEP and EEP tests are complementary and should be carried out
in parallel within experimental and observational
possibilities. This white paper focuses on EEP, and WEP/UFF in particular but this does not preclude the interest of SEP tests like continued and improved lunar laser ranging.

\subsection{The Role of EEP in Theories of Gravitation}
\label{sec:roleEEP}

The EEP is the foundation of all curved space-time or ``metric''
theories of gravitation, including of course GR. It divides
gravitational theories in two classes: metric theories, those that
embody EEP and non-metric theories, those that do not. This
distinction is fundamental, as metric theories describe gravitation as
a geometric phenomenon, namely an effect of curvature of space-time
itself rather than a field over space-time, quite unlike any of the
other known interactions. It might thus appear unnatural to use a
metric theory for gravitation, so different from the formalisms of the
other interactions, and indeed most unification attempts cast doubt on
precisely this hypothesis and thus on the validity of the EEP. Only
experimental tests can settle the question and, in the light of the
above, experimentally testing the EEP becomes truly fundamental.  To
be more precise (see \textit{e.g.} Refs.~\cite{Dicke1964, Thorne1973,
  Will1993}), a metric theory of gravitation is one that satisfies the
following postulates:
\begin{enumerate}
\item Space-time is endowed with a metric tensor $g_{\mu\nu}$, central
  to the metric equation that defines the infinitesimal line element,
  \textit{i.e.} the space-time separation between two events
\begin{equation}\label{metric}
\ud s^2 = g_{\mu\nu}(x^\rho)\ud x^\mu \ud x^\nu\,,
\end{equation}
in some 4-dimensional space-time coordinate system $x^\rho$;
\item The trajectories of freely falling test bodies are geodesics of
  extremal length,
\begin{equation}\label{interval}
\delta \int \ud s = 0\,,
\end{equation}
\textit{i.e.} they depend only on the geometry of space-time, but are
independent of the test body composition;
\item Clocks measure proper time $\tau$ along their trajectory, given
  by
\begin{equation}\label{propertime}
\ud \tau^2 = - \frac{1}{c^2}\ud s^2\,,
\end{equation}
independent of the type of clock used;
\item In local freely falling reference frames, the non-gravitational
  laws of physics (\textit{i.e.} the other three fundamental
  interactions) satisfy the principles of special relativity.
\end{enumerate}
Obviously the above postulates are a direct consequence of the EEP,
for example LLI and LPI are the foundations of points 3 and 4 and WEP
is the basis of point 2. It is important to note that GR is not the
only possible metric theory that satisfies the above
postulates. Indeed, there exist a large number of such theories like
the scalar-tensor Jordan-Brans-Dicke theories~\cite{Jordan1946,
  Brans1961} and their generalizations. These theories differ from GR
in the way that the metric tensor is related to the distribution of
mass-energy through the existence of other fields associated with
gravity (scalar field, vector field, \textit{etc.}).

Theories in which varying non-gravitational coupling constants are
associated with dynamical fields that couple to matter directly are
not metric theories. In such theories, the fine structure constant
$\alpha$ for instance would vary with space and time. Neither, in this
narrow sense, are theories in which one introduces additional fields
(dilatons, moduli) that couple differently to different types of
mass-energy, \textit{e.g.} some versions of Superstring theory. The
fundamental ingredient of all such non-metric theories is
non-universal coupling to gravity of all non-gravitational fields,
\textit{i.e.} the fields of the SM.

Thus experimental tests of the EEP are often viewed as tests of the
universal coupling of gravity (through the metric of space-time
$g_{\mu\nu}$) to all non-gravitational fields of the SM~\cite{Damour2008,Hees2018}. Violations occur when the coupling
is dependent on some attribute of the non-gravitational fields at hand
that may be different for different test bodies, \textit{e.g.}
electromagnetic charge, nuclear charge, total spin, nuclear spin,
quark flavor, Baryon and Lepton numbers, \textit{etc}. Exploring all possibilities
of such anomalous couplings is the fundamental aim of experimental
tests of the EEP. Note also that in any particular experimental
situation, symmetry requires that such anomalous couplings be not only
a function of the composition of the test body, but also of the mass
which is the source of the gravitational field. 

\subsection{Why Would the EEP be Violated?}
\label{sec:whyEEP}

It has already been pointed out that the EEP is in fact rather
unnatural in the sense that it renders gravity so different from other
interactions, because the corresponding universal coupling implies
that gravitation is a geometrical attribute of space-time itself
rather than a field over space-time like all other known
interactions. Einstein himself initially called it the
\textit{hypothesis of equivalence} before elevating it to a
\textit{principle} once it became clear how central it was in the
generalization of special relativity to include gravitation. This
shows how surprising it is in fact that such an hypothesis should be
satisfied at all, let alone down to the uncertainties of present-day
tests. Therefore, rather than asking why the EEP should be violated,
the more natural question to ask is why no violation has been observed
yet. 

Indeed most attempts at quantum gravity and unification theories
lead to a violation of the EEP~\cite{Taylor1988, Damour1994,
  Dimopoulos1996, Antoniadis1998, Rubakov2001, Maartens2010}, which in
general have to be handled by some tuning mechanism in order to make
the theory compatible with existing limits on EEP violation. For
example, in string theory moduli fields need to be rendered massive
(short range)~\cite{Taylor1988} or stabilized by \textit{e.g.}
cosmological considerations~\cite{Damour1994} in order to avoid the
stringent limits already imposed by EEP tests. Similarly M-theory and
Brane-world scenarios using large or compactified extra dimensions
need some mechanism to avoid existing experimental limits from EEP
tests or tests of the inverse square law~\cite{Antoniadis1998,
  Maartens2010, Rubakov2001, Adelberger2009, Antoniadis2011}. Therefore, not only do we expect a violation of EEP at some
level, but the non-observation of such a violation with improving
uncertainty is already one of the major experimental constraints for
the development of new theories in the quest for quantum gravity and
unification. This makes experimental
tests of EEP one of the most essential enterprises
of fundamental physics today.

It is interesting to note that experimental constraints for EEP
violations at low energy are rather closely related to present-day
physics at the very small scale (particle physics) and the very large
scale (cosmology). Notably, the experimental confirmation of the Higgs boson has thus lent strong
credibility to the existence of scalar fields, as the Higgs is the
first fundamental scalar field observed in Nature. It is thus likely
that additional long and/or short range scalar fields exist, as
postulated by many unification theories, and EEP tests are one of the
most promising experimental means for their observation. At the other extreme if such scalar fields are massive they may well constitute the mysterious dark matter (DM) of cosmology. There is no reason for such DM to be universally coupled to SM fields, and it would thus give rise to a violation of the EEP that could be detected by EEP tests~\cite{Hees2018}. Additionally, most models for Dark Energy (DE)
are also based on long-range scalar fields that, when considered in
the context of particle physics, are non-universally coupled to the
fields of the SM~\cite{Khoury2004a, Khoury2004b}. Similar reasoning applies to spin-1 bosonic fields that also may violate the EEP\cite{Fayet2017,fayet19}. As a consequence, one would expect EEP violations from such fields, be it DM and/or DE at some level, which might be low energy experiments, like the ones discussed here. Such a detection would provide a very appealing route towards independent confirmation of DM/DE, making it more tangible than only a hypothesis for otherwise unexplained astronomical observations.

\subsection{EEP in the context of physics today}
\label{sec:context}

\noindent {\it Cosmology:} The big challenge of modern cosmology and
particle physics is to understand the observed ``composition'' of the
Universe {\it i.e.} about 68.3 \% dark energy (DE), 26.8 \% dark matter
(DM), and 4.9 \% baryonic matter~\cite{Ade2013}. These values are
obtained assuming the $\Lambda$-CDM model, in which the vacuum energy density associated to the
cosmological constant is $\rho_\Lambda = \Lambda/8\pi G
\simeq 10^{-47}\,\text{GeV}^4$ ($\simeq\rho_\text{critical}$). On the
other hand, arguments from quantum field theory imply that the vacuum
energy density is the sum of zero point energy of quantum fields with
a cutoff determined by the Planck scale ($m_\text{P} \simeq 1.22
\times 10^{19}\,\text{GeV}$) giving $\rho_\text{vacuum} \simeq
10^{74}\,\text{GeV}^4$, which is about 121 orders of magnitude larger
than the observed value. A lower scale, fixed for example at the QCD
scale, would give $\rho_\text{vacuum} \simeq 10^{-3}\,\text{GeV}^4$
which is still much too large with respect to $\rho_\Lambda$. From a
theoretical point of view, at the moment, there is no explanation as
to why the cosmological constant should assume the correct value at
the scale of the observed Universe. The only argument we can give is
based on the anthropic principle, \textit{i.e.} the idea that much
larger values would not have lead to the formation of stars, planets
and ultimately humans.

Rather than dealing directly with the cosmological constant to explain
the accelerating phase of the present Universe, a number of
alternative approaches and models have been proposed (e.g. \cite{Wetterich1988, Ratra1988, Carroll1998, Khoury2004a, Khoury2004b, Brax2004, Chiba2000, ArmendarizPicon2000, ArmendarizPicon2001, Dvali2000, Kamenshchik2001, Bilic2002, Bento2002, Capozziello2011, Nojiri2007, Caldwell2002}). Many of these models are characterized by the
fact that a scalar or spin-1 field (or more than a single field) is included in the action of
gravity. Additionally the same (or additional) fields may be used to provide the DM required by observations. Again there is no compelling reason why such fields should be coupled universally to the SM fields and thus they would violate the EEP. Hence tests of the EEP are a unique tool to discover the existence of such fields and thus answer one of the most puzzling questions in modern cosmology.\\

\noindent {\it Particle physics:} In the previous paragraph, it already became clear that the difficulties
of GR in cosmology are closely related to those in particle
physics. In particular, in a quantum field theory (like the SM), one would expect that the vacuum energy of
the fundamental fields should be observed in its gravitational
consequences, especially on the large scale of the Universe. However,
there is a huge discrepancy (121 or at least 40 orders of magnitude, see above) between the observed vacuum energy density of
the Universe (dark energy) and the one expected from the SM. This has been considered a major problem in
modern physics, even before the discovery of dark energy when the
``observed'' value of the cosmological constant (or vacuum energy) was
compatible with zero~\cite{Weinberg1989}. And one might argue that
this problem has become even worse since the discovery of the
accelerated expansion of the Universe, and the associated small
\textit{but non-zero} value of $\Lambda$, as now one requires a
mechanism that does not completely ``block'' the gravitational effect
of vacuum energy, but suppresses it by a huge factor, \textit{i.e.}
some extreme fine tuning mechanism is required that is difficult to
imagine.

Another conceptual problem is that the SM requires a number of dimensionless coupling constants to be
put in by hand, which seems somewhat arbitrary and is not very
satisfactory~\cite{Damour2012}. One of the aims of theoretical
developments is then to replace these constants by some dynamical
fields that provide the coupling constants (\textit{e.g.} moduli
fields in string theory, dilaton, \textit{etc.}), similarly to the
Higgs field giving rise to the mass of fundamental particles. As a
consequence the coupling constants become dynamical quantities that
vary in space-time (\textit{e.g.} space-time variation of the fine
structure constant $\alpha$), which necessarily leads to violations of
the EEP. However, the
resulting phenomenological consequences are such that in most
approaches one requires some mechanism to stabilize these fields in
order to be compatible with present-day constraints from EEP
tests~\cite{Taylor1988, Damour1994}. Although no firm predictions
exist, this makes the discovery of the effect of such fields
(\textit{e.g.} EEP violation) a distinct
possibility~\cite{Damour2012}.

Even if one disregards gravity, the SM
still does not address all the fundamental questions: in particular,
whereas it attributes the origin of mass to the Higgs non-vanishing
vacuum value, it does not explain the diversity of the masses of the
fundamental particles, \textit{i.e.} it does not explain the diversity
of the couplings of the matter to the Higgs field. One thus has to go
to theories beyond the SM in order to answer these
questions. Most of these theories make heavy use of scalar fields, the
most notable examples being supersymmetry, which associates a scalar
field to any spin-$\frac{1}{2}$ matter field, string theory and
higher-dimensional theories. Some of these scalar fields may be
extremely light, or even massless, which leads to new types of long
range forces, and thus potential EEP violations, unless these fields
are universally coupled, a difficult property to achieve.\\

\noindent {\it Quantum mechanics and the EEP:} Quantum tests of the Equivalence Principle differ from classical ones
because classical and quantum descriptions of motion are fundamentally
different. In particular, the Universality of Free Fall (or WEP) has a
clear significance in the classical context where it means that
space-time trajectories of test particles do not depend on the
composition of these particles. How UFF/WEP is to be understood in
quantum mechanics is a much more delicate point. The subtlety of
discussions of the EEP in a quantum context is also apparent in the
debate about the comparison of various facets of the EEP, in
particular the UFF and the LPI~\cite{Muller2010, Wolf2011,
  Giulini2012}. More generally, considering quantum phenomena in the
context of gravity poses many conceptual and fundamental difficulties. When comparing classical EEP tests to quantum ones, a number of implicit assumptions are made, like e.g. that quantum mechanics is valid in
the freely falling frame associated with classical test bodies in the
definition of WEP. Indeed, the usual definition of the EEP states that
special relativity holds in the freely falling frame of WEP without
reference to quantum mechanics.\footnote{Recall that relativistic quantum mechanics did not exist at the time of the earliest formulation of the equivalence principle by Einstein.} However, in general the variety of quantum states is much larger than that of classical
ones and it seems therefore plausible that quantum tests may
ultimately be able to see deeper details of couplings between matter
and gravity than classical ones (see \cite{Altschul2015} for a more detailed discussion).

\subsection{Experimental tests of UFF/WEP on ground and in space}

The history of experimental tests of UFF/WEP goes back as far as the Renaissance, and probably beyond. First documented experiments were carried out by Simon Stevin and Galileo Galilei towards the end of the 16th century, followed by Newton, Bessel, E\"otv\"os, Dicke, Braginsky, Adelberger to name only the best known ones. Essentially two different methods were employed, falling objects and torsion balances. On ground, the latter give the best uncertainties \cite{Schlamminger2008} but are ultimately limited by the effect of local gravity gradients. In space the recent CNES/ESA MICROSCOPE mission uses the former and improves on ground experiments by an order of magnitude \cite{Touboul2017}, with another factor 10 improvement expected in the near future, when all data is analysed.

A simple phenomenological figure of merit for all UFF/WEP tests is the E\"otv\"os ratio $\eta_{AB}$ for two test objects
$A$ and $B$ and a specified source mass of the gravitational field:
\begin{equation}\label{etaAB}
\eta_{AB} = 2\,\frac{a_A - a_B}{a_A + a_B}\,,
\end{equation}
where $a_i$ ($i=A,B$) is the gravitational acceleration of object $i$ with respect
to the source mass. Note that for a given experiment the data can be
interpreted with respect to different source masses (see \textit{e.g.}
Ref.~\cite{Schlamminger2008}) with corresponding different results for
$\eta_{AB}$.

Whilst $\eta_{AB}$ is a useful tool for comparing different experiments it cannot account for the diversity of possible underlying theories, e.g. different types of couplings depending on the source and test objects (c.f. the end of section \ref{sec:roleEEP}), or couplings to space-time varying background fields other than local gravity {\it e.g.} \cite{Tasson2011, Hees2018}. Thus, not only best performance in terms of the E\"otv\"os ratio is required, but also a large diversity in terms of test objects and source masses. 

Table \ref{tab:StateArt} presents the state of the art in UFF/WEP tests, separated into different classes as a function of the type of test-masses employed. In particular we distinguish between tests using macroscopic test masses and atom-interferometry (AI) tests that use matter waves in a quantum superposition, possibly condensed to quantum degenerate states (Bose Einstein Condensates) with coherence lengths $\geq \mu$m. The ``game changing'' results of the MICROSCOPE mission demonstrate the potential of going into a quiet and well controlled space environment, with potentially ``infinite'' free fall times.
\begin{table*}[t]
\begin{center}
%\begin{tabular}{|r|r|r|r|r|} 
\begin{tabular}{ccccc}
\hline 
{\bf Class} & {\bf Elements} & {\bf $\eta$} & {\bf Year [ref]} & {\bf Comments} \\
\hline\hline
\multirow{4}{*}{Classical} & Be - Ti & $2\times10^{-13}$ & 2008 \cite{Schlamminger2008} & Torsion balance \\
& Pt - Ti & $1\times10^{-14}$ & 2017 \cite{Touboul2017} & MICROSCOPE first results \\
& Pt - Ti & ($10^{-15}$) & 2019+ & MICROSCOPE full data \\
& \red{$M_A - M_B$} & \red{$10^{-17}$} & \red{2035+} & \red{Adv. MICROSCOPE, macrsocopic masses $M_i$ TBD} \\
\hline
\multirow{3}{*}{Hybrid} & $^{133}$Cs - CC & $7\times10^{-9}$ & 2001 \cite{Peters2001} & \multirow{2}{*}{AI and macroscopic corner cube (CC)}  \\
& $^{87}$Rb - CC & $7\times10^{-9}$ & 2010 \cite{Merlet2010} &  \\
& \red{$At_A - M_B$} & \red{$10^{-17}$} & \red{2035+} & \red{Adv. MICROSCOPE, atomic species $At_A$ TBD} \\
\hline
\multirow{5}{*}{Quantum} & $^{39}$K - $^{87}$Rb & $5\times10^{-7}$ & 2014 \cite{Schlippert2014} & different elements \\
& $^{87}$Sr - $^{88}$Sr & $2\times10^{-7}$ & 2014 \cite{Tarallo2014} & same element, fermion vs. boson  \\
& $^{85}$Rb - $^{87}$Rb & $3\times10^{-8}$ & 2015 \cite{Zhou2015} & same element, different isotopes  \\
& $^{85}$Rb - $^{87}$Rb & ($10^{-13}$) & 2020+ \cite{Overstreet2018} & \multirow{2}{*}{$\geq$ 10 m towers} \\
& $^{170}$Yb - $^{87}$Rb & ($10^{-13}$) & 2020+ \cite{Hartwig2015} & \\
& \red{$^{41}$K - $^{87}$Rb} & \red{$10^{-17}$} & \red{2035+} & \red{Atom Interferometry mission} \\
\hline
Antimatter & $\overline{\rm H}$ - H & ($10^{-2}$) & 2020+ \cite{Doser2010,Perez2012} & under construction at CERN \\
\hline
\end{tabular}
\caption{State of the art in UFF/WEP tests. Numbers in brackets are results expected in the near future. In red the performances aimed for in this white-paper.}\label{tab:StateArt}
\end{center}
\end{table*}

\begin{figure}[!tbp]
  \centering
  \begin{minipage}[b]{0.49\textwidth}
    \includegraphics[width=\textwidth]{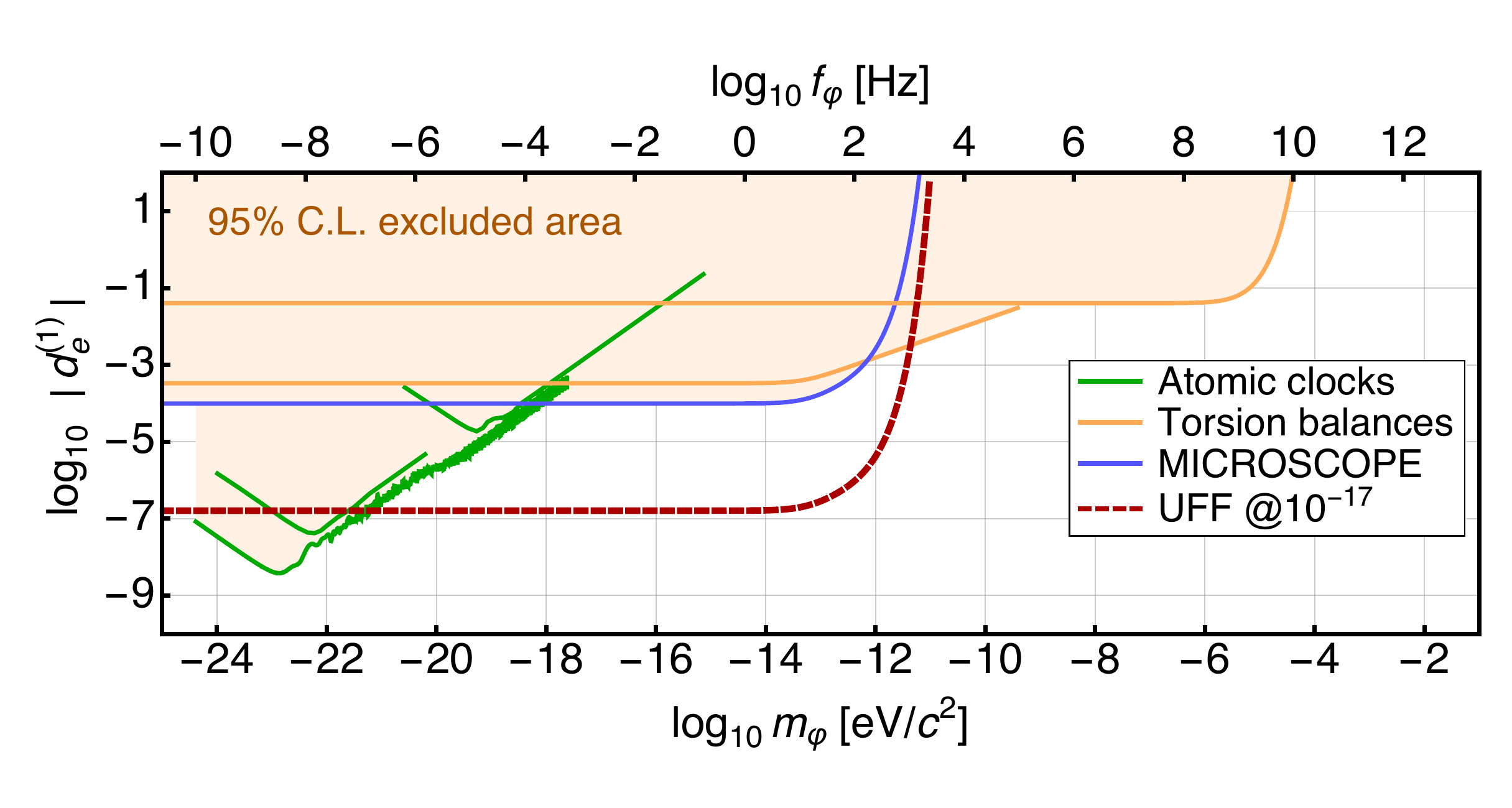}
    \caption{Constraints for scalar DM, linear coupling.} \label{fig:Hees1}
  \end{minipage}
  \hfill
  \begin{minipage}[b]{0.49\textwidth}
    \includegraphics[width=\textwidth]{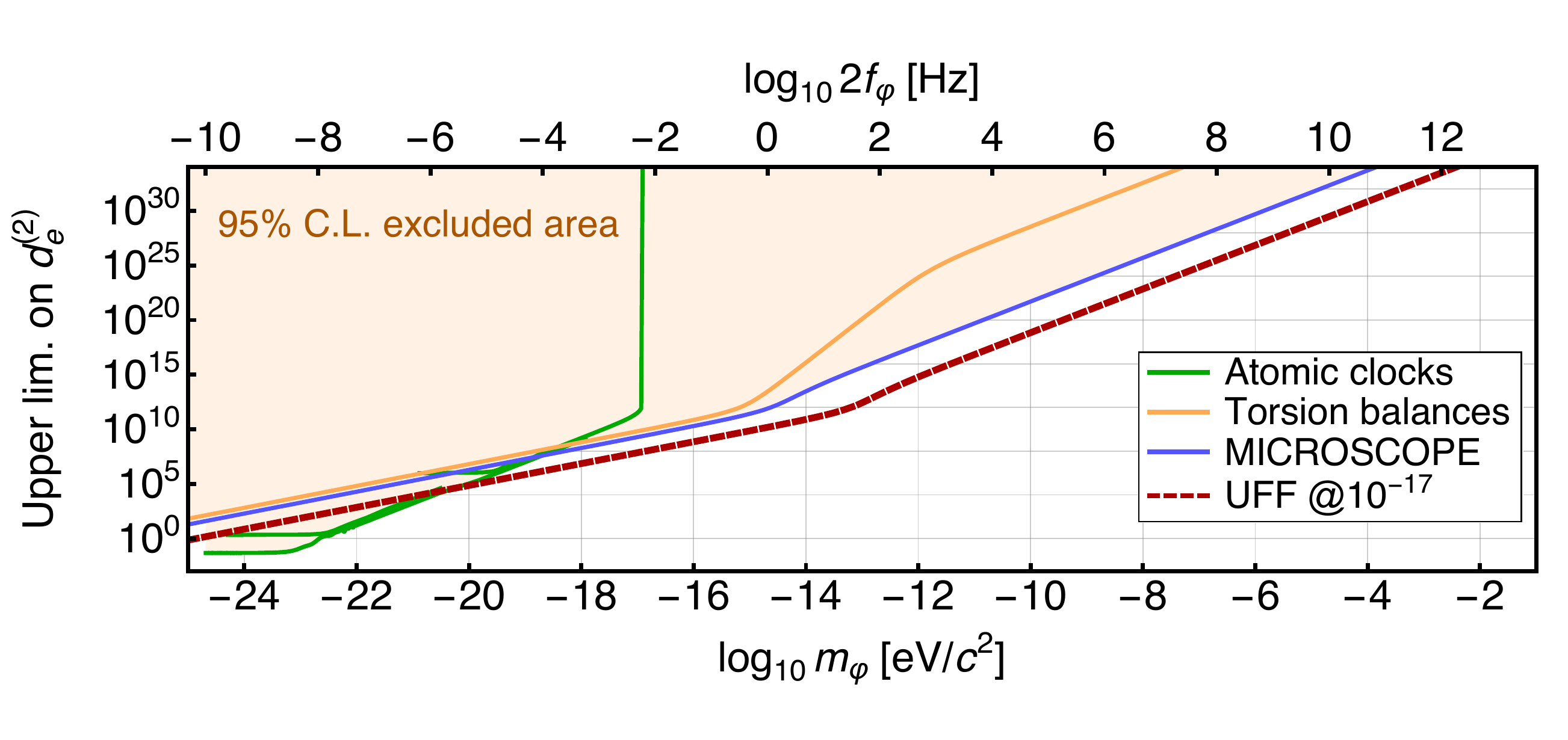}
    \caption{Constraints for scalar DM, quadratic coupling.} \label{fig:Hees2}
  \end{minipage}
\end{figure}

As an example of a more fundamental theory that is constrained by UFF/WEP tests, and to demonstrate the link to the enigma posed by dark matter, we also show the analysis of different types of experiments in a theory where dark matter is represented by a massive but light scalar field that is non universally coupled to the SM (see \cite{Damour2010,Arvanitaki2015,Stadnik2015,Hees2018} for details). In such a theory objects of different composition fall differently and clocks using different atomic transitions run at different frequencies. Depending on the type of coupling involved those differences can have very specific spatio-temporal signatures that can be searched for in the data. The basic interaction Lagrangian in such a theory is written
\begin{equation}\label{eq:Lint_quad}
 \mathcal L_\textrm{int}^{(k)} =	\frac{\varphi^k(t,\bm x)}{2}  \Bigg[\frac{d^{(k)}_e}{4e^2}F^2-\frac{d_g^{(k)}\beta_3}{2g_3}\left(F^A\right)^2 -\sum_{i=e,u,d}\Big(d_{m_i}^{(k)}+\gamma_{m_i}d_g^{(k)}\Big)m_i\bar\psi_i\psi_i\Bigg] \, , 
\end{equation}
with $F_{\mu\nu}$ the standard electromagnetic Faraday tensor, $e$ the electric charge of the electron, $F^A_{\mu\nu}$ the gluon strength tensor, $g_3$ the QCD gauge coupling, $\beta_3$ the $\beta$ function for the running of $g_3$, $m_i$ the mass of the SM fermions, $\gamma_{m_j}$ the anomalous dimension giving the energy running of the masses of the QCD coupled fermions and $\psi_i$ the fermion spinors.  The constants $d_j^{(k)}$ characterize the interaction between the scalar field $\varphi$ and the different matter sectors, with $k=1,2$ corresponding to the simple cases of linear or quadratic coupling, and are determined by experiment. Figures \ref{fig:Hees1}, \ref{fig:Hees2} show constraints on coupling to electromagnetism ($d_e$) when assuming that all of the DM is made up of $\varphi^k(t,\bm x)$. The predominance of UFF/WEP tests is manifest, especially at larger masses of the DM field. More importantly, in the present context, the MICROSCOPE space mission has improved previous knowledge by one to ten orders of magnitude, depending on the assumed coupling and DM mass, and the projects discussed in this white paper are expected to improve on that by another 3-4 orders of magnitude.

\subsection{Secondary science objectives} \label{sec:secondary}

\noindent {\it Other aspects of the EEP:} When testing the WEP/UFF one generally also tests the other aspects of the EPP, {\it i.e.} Local Lorentz and Position Invariance (LLI and LPI). Additionally, LLI is closely related to CPT invariance. The exact link between the different tests is model dependent. For example, a recent analysis of MICROSCOPE data in a very general LLI-violating framework called the SME (Standard Model Extension) gives large improvements on the constraints on four SME coefficients that govern a possible LLI violation in the coupling between gravity and the SM fields \cite{pihan17, Pihan2019}. Depending on the other instruments on board and the chosen orbit one may also carry out tests of LPI (via the gravitational redshift) in the field of the Earth (with an on board clock) or in the field of the Sun and Moon (with only a time/frequency comparison system on board, but no clock), all of which can significantly enhance present knowledge \cite{Altschul2015,YellowBook,Wolf2016}. Such experiments will naturally benefit from the heritage of near future gravitational redshift tests like ACES \cite{Cacciapuoti2009,Savalle2019} and SOC/I-SOC \cite{Bongs2015,Origlia2016,Schiller2017}. \\

\noindent {\it Time/frequency metrology:} Closely related to the LPI tests is the possibility of comparing ground clocks over intercontinental distances without degrading their performance. Presently such comparisons are done using space techniques (navigation and telecom satellites) but their uncertainties are two or more orders of magnitude larger than those of the clocks themselves, and thus hamper their use in applications ranging from fundamental physics to geodesy and international time scales (TAI). One way around that problem is to add high performance time/frequency links to satellites like STE-QUEST \cite{YellowBook,Altschul2015}, as already planned for ACES \cite{Cacciapuoti2009,Meynadier2018} and SOC/I-SOC \cite{Bongs2015,Origlia2016,Schiller2017}. Flight models of a high-performance two-way time/frequency microwave link and of a single-photon time transfer link have been developed and will be flown on the ACES mission. An improved microwave link (100 times lower noise, "HERO") is currently in the breadboarding phase within an ESA-funded industrial project. The technology for an improved single-photon link (8 times lower uncertainty in time transfer) has also been developed and is ready to be turned into a flight model \cite{Panek2010,Prochazka2013}. \\

\noindent {\it Geodesy and Reference Frames:} Any mission with high performance accelerometers on board has the capacity for inertial navigation and the determination of a purely gravitational trajectory {\it i.e.} purely geodesic motion \cite{Touboul2017,Armano2019}. If furthermore one or several orbit determination methods are available ({\it e.g.} GNSS, DORIS, SLR) the mission provides a means of mapping local space-time and its metric, {\it i.e.} the local gravitational field. This is of particular interest in terrestrial orbit as such a mission can contribute to the determination of the geopotential and its variations, with applications of prime importance in fields as diverse as hydrology or earthquake precursory signals \cite{Panet2018}. If additionally time/frequency comparison methods are included it opens the way to point-measurements of the geopotential at the location of ground clocks with sub-cm ($< 0.1$~m$^2$s$^{-2}$) uncertainty when using today's best ground clocks, a method known as chronometric geodesy \cite{Lion2017,Denker2017,Mehlstaeubler2018}. Finally, akin to the E-GRASP mission proposal \cite{Biancale2016}, such a mission can serve as a common reference point in space for different geodetic techniques thus unifying Terrestrial reference frames at the mm level, which is critical when {\it e.g.} trying to measure sea-level changes of the order of a mm/year. 

\subsection{Summary} \label{sec:summary}

The EEP is at the heart of modern physics and closely intertwined with some of the most fundamental questions of gravitation, particle physics and cosmology. UFF/WEP tests are a unique opportunity to find answers to some of those questions, with the potential for a major discovery when improving present performance by two or more orders of magnitude. To do so, only space offers the required quiet and well-controlled environment together with long free fall times, both of which are indispensable to further advance the field, as convincingly demonstrated by the MICROSCOPE mission. In terms of space-technology Europe has an undisputed world-wide lead in this endeavour through the MICROSCOPE and LISA-Pathfinder missions, both of which have demonstrated unrivalled performance in drag-free control and accelerometry \cite{Touboul2017,Armano2018,Armano2019}. It is now time to build on that heritage for exciting science of the next decades.   

\section{Technological Impact} \label{sec:techno}

In many cases the technologies pioneered in fundamental physics missions significantly improve and continue to have a positive effect on more applied fields. Missions testing the EEP require high quality drag-free motion which can be achieved using highly sensitive and stable accelerometers \cite{Touboul2017, Armano2018, Armano2019}. When also testing LPI (gravitational redshift) they require high precision time/frequency metrology (clocks, time transfer methods).

Highly sensitive and stable accelerometers are also required in missions exploring the gravitational field, in particular in geodesy missions. The GRACE \cite{Tapley2004,Flury2008} accelerometer performance was $\sim 10^{-10} \; \text{m}/\text{s}^2 /\sqrt{\text{Hz}}$, for GOCE \cite{Bock2011} one had $\sim 10^{-12} \; \text{m}/\text{s}^2 /\sqrt{\text{Hz}}$. GOCE had to be more sensitive on shorter time scales (resulting in better performance on smaller spatial scales) than GRACE which has the best performance on larger spatial scales. The MICROSCOPE accelerometers have a performance of $\sim 10^{-11} \; \text{m}/\text{s}^2 /\sqrt{\text{Hz}}$ with LISA-Pathfinder reaching as low as $\sim 10^{-15} \; \text{m}/\text{s}^2 /\sqrt{\text{Hz}}$. Similarly, several recent and ongoing (ESA, CNES, DLR) studies are exploring the potential of cold-atom inertial sensors in geodesy missions and related applications. As an example, a very recent publication \cite{Abich2019} shows nm performance of the laser ranging interferometer (LRI) on GRACE-FO. In order to to be able to turn that into useful geodetic information, accelerometers are required, which are one to two orders of magnitude more precise than those currently available. Also for astrometry missions or a VLBI constellation mission in space a precise knowledge of the motion and, thus, a stable and precise inertial sensor is needed.

Highly stable space clocks and time/frequency transfer methods, in particular optical ones, have applications in navigation, intercontinental clock comparisons and international time scales, broadband telecommunications and chronometric geodesy, i.e. determination of Earth potential differences between particular locations at the cm and sub-cm level \cite{Lion2017, Denker2017, Mehlstaeubler2018}. 

\section{An Atom Interferometric space test of UFF/WEP}
\subsection{Introduction and objectives}

The coherent manipulation of cold atoms with electromagnetic fields is key to new types of sensors with various metrological applications. Indeed, time and frequency are today's best realized physical units, thanks to atomic clocks based on optical and microwave transitions.
%As of today, state-of-the art clocks reach $6\times 10^{-18}$ in fractional uncertainty and $\sigma_y(\tau) = 3\times 10^{-16} \tau^{-1/2}$ in terms of stability, expressed as Allan deviation.
Moreover, freely falling atoms constitute excellent test masses, hence allowing to infer inertial quantities through interferometric measurements.
In particular, their long-term stability and high accuracy renders atomic gyroscopes \cite{Savoie2018} and accelerometers \cite{Freier2016} exquisite tool for navigational, geodesic and fundamental \cite{Schlippert2014,Rosi2017,Zhou2015,Overstreet2017,Parker2018} applications.

A concurrent operation of two such accelerometers with different atomic species provides a new pathway to tests of the UFF. These experiments extend the range of test pairs significantly to previously inaccessible species and hence prove invaluable to explore many facets of different violation scenarios such as the SME \cite{Hohensee2013}. Moreover, phenomena exclusive to quantum systems, such as coupling of gravity to spin \cite{Tarallo2014} or to superpositions of electronic states \cite{Rosi2017}, provide unique insight into the interface  of gravity and quantum mechanics. It is important to stress here that this class of experiments is truly quantum in nature, in particular: 
\begin{itemize}
    \item The observable is the phase difference of interfering matter waves in a coherent superposition;
    \item The coherent superposition is well separated spatially by $>$ 10 cm, more than 3 orders of magnitude larger than the size of the individual wave packets;
    \item The coherence length of the atoms is of the order of a micron, many orders of magnitude larger than the de Broglie wavelength of the macroscopic test masses ($10^{-27}$ m or less).
\end{itemize}

Finally, on the technical side, the properties of atoms and their interaction with the environment can be controlled to high accuracy, which allows to realize test masses of highest isotopic purity and to mitigate systematics.  
In this sense, quantum mechanics allow for several unique advantages. For instance, the atoms can be condensed to a quantum degenerate state (Bose Einstein Condensates, BEC), which has very favorable phase-space properties such as ultra-low expansion rates. Also, various malicious effects couple to the displacement of the two test masses upon release. With matter waves, it is possible to truly overlap the two species and image them simultaneously in situ. As of today, quantum tests of the UFF have reached uncertainties in the range of $10^{-7}-10^{-8}$ in the \Eotvos ratio \cite{Schlippert2014,Zhou2015,Barrett2016}, with the prospect of reaching uncertainties beyond $10^{-13}$ in long-baseline setups \cite{Overstreet2017,Hartwig2015}. The sensitivity of an atom interferometer to acceleration scales quadratically with the free fall time of the atoms, which eventually limits the possible performance of  ground-based experiments. As a consequence, space-borne missions with in principle unlimited drift times are the natural ambition for highly accurate quantum tests of the UFF \cite{Williams2016,Aguilera2014}. Space offers further exceptional advantages for atom optics, such as the possibility for symmetric beam splitting and release from shallow traps, which inherently suppresses noise sources related to laser phase noise and atomic ensemble temperature. Furthermore, temporarily varying configurations of a satellite with respect to the gravitational field of the Earth allow for modulations that distinguish a potential UFF violation signal from systematic uncertainties. 

In the following, we will discuss a mission concept for such a quantum test of the UFF on a circular, low-Earth-orbit, which allows for a target uncertainty in the \Eotvos ratio below $10^{-17}$ as primary mission goal. This implies a three orders of magnitude improvement over current limits. Thanks to discovery of new techniques that attenuate various systematic effects and due to the rapid developments concerning space-maturity of quantum systems, the presented mission provides a promising concept for a quantum test of the UFF with unprecedented accuracy, based on state-of-the-art technology. Note that this is a only an example mission scenario for the purposes of this white-paper, a detailed trade off study in terms of primary and secondary science objectives (c.f. sect. \ref{sec:secondary}) and mass, consumption, cost etc... must be carried out for a complete mission proposal.

\subsection{Atom interferometric test of the UFF}

Atom interferometry exploits the wave nature of matter to infer metrological quantities through interference. To this end, freely falling matter waves are subject to a series of light pulses, which serve as beam splitters and mirrors in close analogy to optical Mach-Zehnder interferometers. Through a stimulated two-photon process, such a light pulse can transfer momentum to an atom and imprints a position dependent phase. A first beam splitter puts the atoms into a superposition of two motional states, which travel along different trajectories before being redirected by a mirror pulse and finally recombined by another beam splitter. The two output ports of the interferometer differ in momentum, and their relative population depends on the phase difference accumulated between the two branches. Since at each light pulse, the position of the atoms is referenced to the light field, this phase difference is indicative of the free fall acceleration $a$ of the matter waves with respect to the apparatus. To first order, the phase is $\Delta \phi = K a T^2$, where $K$ is the effective wave number quantifying the momentum transferred at each pulse and $T$ is the pulse separation time. In a differential measurement with two species $A$ and $B$, the differential acceleration uncertainty per experimental cycle
\begin{equation}
        \sigma_{\Delta a} = \left[\left(\frac{1}{C_A K_A T_A^2 \sqrt{N_A}}\right)^2 +\left(\frac{1}{C_B K_B T_B^2 \sqrt{N_B}}\right)^2\right]^{1/2}
    \end{equation}
is limited by the quantum-projection noise (shot noise), given by the number $N$ of atoms contributing to the signal. The contrast $C$ accounts for the visibility of the interference fringes.  Typically, a retro-reflective setup is employed, such that the same mirror serves as a reference for both interferometers, which are operated simultaneously. This leads to common mode rejection for various systematics and noise sources, where the suppression factor depends on the choice of the atomic species. 

Ultimately, the experiment proposed here monitors the motion of two atomic wave packets with initially superposed centers. It can be interpreted as a test of classical general relativity coupled to a Klein-Gordon field in a non-relativistic limit or, equivalently, a Schr\"odinger equation with an external gravitational potential.
The sensitivity to violations of the UFF is quantified by the \Eotvos ratio $\sigma_{\Delta a}/g$ and suggests operation on a low-earth-orbit.

\subsection{Operation mode}

The shot-noise limited uncertainty in the \Eotvos ratio displays the maximal achievable sensitivity to a potential violation signal possible with such sensors, given that systematic and stochastic errors can be kept below this level. As white noise, it may be averaged down with many repeated cycles.
In the following, we will consider a space-borne mission on a circular orbit, where the satellite is kept inertial with respect to distant stars. For the determination of the \Eotvos ratio, the integration of the signal needs to take into account the varying projection of the gravitational acceleration $g$ onto the sensitive axis \cite{Schubert2013}, such that the averaging over $n$ measurements reads
\begin{equation}
\sigma_\eta = \frac {1} {\sqrt n}\sqrt{\frac{1}{n-1}\sum^n_{j=1} \left(\frac{\sigma_{\Delta a}}{g(t_j)}\right)^2}.
\end{equation}
The number of beneficial measurements per orbit is limited, since the local value of $g$ becomes too small for certain orbital positions. For an inertial satellite on a circular orbit with orbital frequency $\Omega$, it can be written as $g(t_j) = g_0 \cos(j \Omega T_c)$.

Aiming for a target uncertainty of $\sigma_{\eta} \leq 10^{-17}$ suggests parameters as presented in Table \ref{tab:params}. We assume a moderate beam splitting order of $2$ in order to keep the spatial extent of the interferometers below one meter. Moreover, we suppose typical atomic numbers and cycle time for the generation and engineering of BECs. Assuming that $\SI{10}{s}$ are required for the atomic source preparation, followed by an interferometer of $2T=\SI{40}{s}$ duration, the stated cycle time requires an interleaved operation of 5 concurrent interferometers. The contrast can be assumed to be near unity, since major sources of contrast loss, such as gravity gradients, can be mitigated as will be outlined later. Given an altitude of $a=\SI{700}{km}$ height and a cycle time of $T_c = \SI{10}{s}$, a maximum of 356 measurements per orbit allows to integrate the shot-noise limited \Eotvos ratio to $8.8\times 10^{-16}$ after one orbit, such that a total of $\tau = \SI{18}{months}$ of integration are required to reach $\sigma_{\eta} \leq 10^{-17}$. \\

\begin{table}
        \centering
        \begin{tabular}{l|c}
           	
           	\multicolumn{2}{c}{Parameters}	\\
			\hline \\ [-1.5ex]
            Atom number $N$ &  $10^6$\\
            Effective wave number $K$ & \\
            \hspace{1cm} Rb & $8\pi/(\SI{780}{nm})$ \\
            \hspace{1cm} K & $8\pi/(\SI{767}{nm})$\\
            Free evolution time $T$ & $\SI{20}{s}$\\
            Cycle time $T_c$ & $\SI{10}{s}$\\
            Simultaneous interferometers & 5 \\
            Contrast $C$ & 1\\
            Orbit \\
            \hspace{1cm} Semi major axis $a$ & $\SI{700}{km}$\\
            \hspace{1cm} Ellipticity $e$ & $\leq 10^{-4}$\\[2mm]        
    \hline \\[-1.5ex]
%            Single shot diff. acc. sensitivity & $\SI{1.09}{\times 10^{-13}~m/s^2}$\\
            Single shot diff. acc. sensitivity & $1.09 \times 10^{-13}$~m/s$^2$\\
%            Integration over one orbit & $\SI{8.8}{\times 10^{-16}~m/s^2}$ \\
            Integration over one orbit & $8.8 \times 10^{-16}$~m/s$^2$ \\
            Integration time to $\delta \eta = 10^{-17}$ & $\SI{18}{months}$ \\[2mm]
        \end{tabular}
        \caption{Parameters for a quantum test of the UFF targeting $\delta \eta \leq 10^{-17}$.}
        \label{tab:params}
    \end{table}

\subsection{Mission requirements}

Any spurious differential acceleration between the two species can, a priori, not be distinguished from a potential UFF violation signal. Consequently, random acceleration contributions need to be kept below shot-noise. All systematic error sources have to be controlled at a level better than the target inaccuracy of $10^{-17}$, or be modulated at other frequencies than the local projection of $g$. In general, one can  decompose the differential acceleration into its frequency components,
\begin{equation}\label{eq:acc_freq}
    \Delta a = \delta a\cos(\omega_0 t)+\Delta a_\text{const}+\sum_{j=0} \Delta a_\text{sys}^j \cos(\omega_j t),
\end{equation}
where $\delta a$ is the potential violation signal that is to be detected, $\Delta a_\text{sys}^j$ a systematic acceleration contribution at frequency $\omega_j$ and $\Delta a_\text{const}$ comprises all non-modulated terms \cite{Williams2016}. Demodulation of the signal frequency $\omega_0$,  at which a possible violation signal is expected, averages all other frequency components down,
\begin{equation}\label{eq:avg}
 \frac{2}{\tau} \int_0^\tau \Delta a \cos(\omega_0 t) dt  \leq \left(\delta a+\Delta a_\text{sys}^0 \right) + \frac{2}{\tau \omega_0}\left(\frac{\delta a}{2}+\left|\Delta a_\text{const}\right|+\frac 4 3\sum_{j=1}\left|\Delta a_\text{sys}^{(j)}\right|\right),
\end{equation}
where $\tau$ is the duration of integration. This is a pessimistic upper bound, since for appropriate choices of $\tau$, the integral over certain frequency components is trivial. The key insight is that the violation signal is demodulated to dc, while all systematic contributions are averaged down at a rate inversely proportional to $\omega_0$. This fact is, for example, employed in \cite{Touboul2017}, where the satellite is additionally spun for an improved integration rate. We, however, consider a mission in which the satellite is kept inertial with respect to distant stars, such that $\omega_0$ corresponds to the orbital frequency. Obviously, differential acceleration contributions $\Delta a_\text{sys}^0$ modulated at $\omega_0$ can not be discriminated from a potential violation signal with this technique, and therefore have to be well-controlled. \\

\paragraph{Gravity gradients and rotations} Any deviation $\Gamma = \partial_r g$ from a uniform gravitational potential as well as rotation of the apparatus (rigidly attached to the satellite) couple to the phase-space distribution of the atoms. This gives rise to additional accelerations proportional to the initial displacement $\Delta r$ and differential velocity $\Delta v$ of the two atomic clouds upon release, which puts strict constraints on their preparation. In fact, equalizing the release of the test masses is a challenge that is common to all types of free fall tests of the UFF \cite{Blaser2001}. 
However, compensating for the gravity gradient induced acceleration terms by introducing additional, experimentally controllable shifts \cite{Roura2017} allows to alleviate the requirements on the atomic source design and mission duration significantly. To this end, the frequency of the mirror pulse is shifted by a few hundred MHz in order to change the effective momentum by $\Delta K = K \Gamma T^2 /2$ in a ground-based experiment \cite{Overstreet2017,Damico2017}. In order to account for the varying projection of the gravity gradients onto the sensitive axis of the interferometer in a (non-nadir) space mission, the laser realizing the mirror pulse has to be periodically tilted about $400 \mu$rad and shifted in frequency by $\SI{150}{GHz}$ for the parameters stated in Table \ref{tab:params}. As a consequence, the overlap of the two atomic clouds only needs be realized within $\Delta r = \SI{100}{nm}$ in position and $\Delta v = \SI{10}{nm/s}$ in velocity, respectively. This has significant implications for the time required to characterize systematics: The center-of-mass position of an atomic ensemble of size $\sigma_r$ can be determined within an accuracy $\sigma_r/\sqrt{\nu}$, given by the statistical distribution of the atoms (similar for the velocity). In order to verify that the target spatial overlap accuracy is given, the required number $\nu$ of images made of that cloud can hence be very large, which has been invoked as a major argument against previous proposals \cite{Aguilera2014} of the UFF with space-borne quantum tests \cite{Nobili2016}. Although the target accuracy of the present proposal is two orders of magnitude more ambitious than the previous, the requirements on the preparation of the atoms are relaxed by two orders of magnitude (corresponding to four orders less verification time) thanks to this technique.

Similarly, the effect of large rotation rates $\Omega_m$ as for a spinning satellite can be compensated by counter-rotating the retro-reflex mirror by an angle $\Omega_m T$ between two subsequent pulses \cite{Lan2012}. This is, however, not required in the present mission scenario employing an inertial configuration.
The allowed spurious rotations of the satellite are constrained by the center of mass velocity jitter of the atomic clouds, which is determined by their temperature via $\sqrt{k_B T/m}/\sqrt{N}$, where $k_B$ is the Boltzmann constant. Supposing an effective ensemble temperature of $T=\SI{10}{pK}$ requires the satellite rotation to be controlled at the nrad/s level. Since these temperatures already comply with state of the art \cite{Kovachy2015,Rudolph2016} and paths to even lower energies exist, these constraints will be more relaxed in the future. Recently, the LISA Pathfinder platform already demonstrated excellent satellite attitude control in the range of 0.1 - 10~nrad/s$^2\sqrt{\text{Hz}}$ \cite{Armano2019}.

\paragraph{Spurious linear accelerations}
Vibrational noise, leading to a random acceleration of the mirror, can not be distinguished from inertial acceleration. However, this effect can be suppressed by matching the frequency response of the two interferometers \cite{Cheinet2008}.
Also, correlation to classical acceleration sensors allows for post-correction and fringe reconstruction even in very noisy environments \cite{Barrett2016,Langlois2017,Richardson2019}, allowing for a high dynamic range. Currently, the performance of these hybrid sensors is limited by the performance of the classical acceleration sensor, calling for improvement on that end.
A vibrational background power spectral density of less than $10^{-9}$~m/s$^2/\sqrt{\text{Hz}}$ for low frequencies has already been demonstrated in GRACE \cite{Flury2008} and LISA Pathfinder \cite{Armano2019}.

\paragraph{Beam splitting laser linewidth} 
In the retro-reflective setup, one of the two light-fields realizing the two-photon process in the atom is reflected at the mirror, while the other is not. The resulting time delay between the two lasers results in a sensitivity to frequency jitter of the beam splitter lasers. Given the spatial extent of the interferometers, this time delay is about $\SI{3}{ns}$ ($\SI{5}{ns}$) for rubidium (potassium). Following \cite{LeGouet2007}, a Lorentzian linewidth of $\SI{20}{kHz}$ ($\SI{1}{kHz}$) integrated over the beam splitter pulse duration of 100~$\mu$s gives rise to a noise contribution per experimental cycle below $10^{-13}$~m/s$^2$, which is in line with the requirements.

\paragraph{Wave front aberrations}
As outlined above, the phase fronts of the light pulses serve as reference to trace the free fall of the atoms. However, any deviations from a planar wave front lead to shifts that vary over the spatial extent of the atomic cloud \cite{LouchetChauvet2011}. This gives rise to a spurious acceleration $\Delta a_{wf} = \sigma_v^2/R$, where $R$ quantifies the curvature of the mirror and $\sigma_v^2 = k_B T/m$ the effective expansion rate of the atomic ensemble. If the shot-to-shot fluctuation in temperature $T$ and wave front can both be constrained below $<10\%$, respectively, this effect can be assumed to be sufficiently constant to be suppressed in the signal demodulation.  

\paragraph{Magnetic fields} In order to exclude a linear Zeeman shift, both interferometers are operated with atoms in the magnetic $m_f=0$ substate. However, the quadratic Zeeman shift leads to a differential acceleration $\Delta a_B = h B_0 \delta B (k_{Rb}/m_{Rb}-k_K/m_K)$, with $k_{Rb} = \SI{575.15}{Hz/G^2}$ and $k_K = \SI{1294}{Hz/G^2}$ In order for magnetic field fluctation induced acceleration noise to stay below shot-noise, the offset magnetic field $B_0$ needs to be stabilized at $\SI{1}{mG}$, while the magnetic field gradients $\delta B$ have to be controlled at the level of 1$\mu$~G/m.

\paragraph{Mean field}
Since the mean field energy scales with the density, any fluctuation in the number of atoms between the two interferometer arms introduces additional noise. Consequently, given a shot-noise limited accuracy of 0.1\% for the first beam splitting ratio and an isotropic expansion corresponding to $\SI{10}{pK}$, the required cloud size is $\sigma = \SI{6}{mm}$ in order to constrain this noise source within a few $\SI{0.1}{mrad}$ \cite{Debs2011}.

\paragraph{Blackbody radiation} Thermal radiation causes an ac-Stark shift of the energy levels in an atomic system. It has recently been pointed out \cite{Haslinger2018}, that this effect is not only of importance in atomic clocks \cite{Nicholson2015} but that temperature gradients, leading to ac-Stark shift gradients, act as a force that accelerates the atoms. For an order of magnitude estimation, the experiment is assumed to be performed inside a thin cylindrical vacuum chamber with a linear temperature gradient $\delta T_{exp}$, which leads to an acceleration $\Delta a_{BBR} = 2 \alpha_0 \sigma  \partial_z (T_{exp}+\delta T_{exp}z)^4 /(m c \epsilon_0)$, where $\alpha_0$, $\sigma$ and $\epsilon_0$ denote the static polarizability of the atom, the Stephan-Boltzmann constant and the vacuum permittivity, respectively.  In order to constrain this effect below shot noise,  the shot-to-shot variations in the temperature and its  gradient need to to be constrained within $\SI{0.5}{K}$ and $\SI{0.5}{mK/m}$, respectively, assuming a chamber temperature $T_{exp} = \SI{300}{K}$ and gradient $\SI{0.1}{K/m}$.

\subsection{Payload performance and ground segment}

The payload performs repeated measurement cycles based on predefined sequences.
In nominal operation, the initial step is the production of a dual species Bose-Einstein condensate of $^{87}$Rb and $^{41}$K starting from an on-chip 3D magneto-optical trap loaded by a cold beam from a 2D magneto-optical trap, molasses cooling, evaporation inside in a magnetic chip trap, transfer to and condensation inside an optical dipole trap.
Herein, a Feshbach field of about 70\,G~\cite{Thalhammer2008,Ferrari2002} ensures miscibility of the two species with $10^6$\,atoms each and is switched off later after a delta-kick collimation~\cite{Rudolph2016,Kovachy2015,Muntinga2013}, reducing residual expansion rates to below 100\,{\textmu}m/s.
Subsequently, the atoms are transferred to magnetically less sensitive states.
Generation and preparation of the atoms is expected to take 10\,s.
The interferometry consists of three laser pulses with lengths of about 100\,{\textmu}s simultaneously applied to both species and separated by $T=20\,\mathrm{s}$ driving double Raman transitions~\cite{Leveque2009}.
Each atom-light interaction imprints a phase onto the atoms, depending on the distance of the atoms to the mirror which retro reflects the light fields for the pulses.
The total of the phases determines the population of the two output ports which are subsequently read out with a spatially resolved detection.
From the knowledge of the pulse timings, their wave numbers, and direct signal extraction~\cite{Schlippert2014} or correlation with an additional accelerometer for noise rejection~\cite{Barrett2016}, the differential acceleration of the two species is estimated~\cite{Hogan2008,Cheinet2008,Bongs2006,Borde2004}.
The duration of a single experiment run is estimated to 50\,s, implying five concurrent~\cite{Savoie2018} experiments in nominal operation.
Adjusting the wave length of the three laser pulses for interferometry~\cite{Roura2017} and tilting the retro reflection mirror~\cite{Freier2016,Lan2012,Hogan2008} according to known gravity gradients and rotations suppresses spurious phase terms arising from initially imperfect overlap of the two species~\cite{Aguilera2014}.\\
The payload consists of three main parts~\cite{Schuldt2015}, the electronics package, the laser system, and the physics package.
The electronics package houses the data management unit which executes the sequences, stores, and processes data, as well as frequency chains, drivers, and controllers for lasers, switching elements, magnetic field generation, CCD cameras, RF and microwave antennae, and other sensors inside the other two parts.
Lasers and optical benches generating the light fields for cooling, preparation, interferometer pulses, detection of $^{87}$Rb at 780\,nm and $^{41}$K at 767\,nm, as well as the dipole trap at 1560\,nm are located in the laser system.
The design is based on high power laser diodes and telecom technology and supports the detuning for the gravity gradient compensation technique.
Inside the physics package, an ultra-high vacuum chamber accomodates the 1\,m baseline of the atom interferometer.
It features an atom chip for rapid generation of cold atoms~\cite{Rudolph2015}, and a retro reflection mirror on a tip-tilt mount inside the vacuum vessel, is connected to an additional smaller chamber supplying a beam of cold atoms for loading the 3D magneto-optical traps, and to an ion getter pump, the latter maintaining the vacuum.
Coils for the generation of offset fields and the Feshbach field of about 70\,G, optics for beam shaping, CCD cameras, an accelerometer, other sensors as photo diodes and thermistors, and structural elements are attached externally.
A \textmu-metal shield surrounds the chamber and mounted elements excluding the pump to suppress external magnetic stray fields.\\
During science commissioning, parameters are partly autonomously optimised, partly adjusted by the ground station depending on the downlinked data, and subsequently verified by uploaded and autonomously executed sequences.
Nominal operation foresees autonomous execution of experimental sequences predefined and uploaded by the ground station, and downlinking of science and housekeeping data to the ground station for storage and further processing.

\subsection{Heritage and development activities}

A multitude of national, European and international programs are a prime heritage to a satellite atom interferometric test of the Equivalence Principle. 

On the studies side, space mission concepts involving cold atoms were presented several decades ago during the HYPER mission proposal~\cite{Jentsch2004}. The ESA project SAI has developed a transportable atom interferometer for ground tests~\cite{Sorrentino2010}. The Q-WEP study was concerned with a UFF test on board of the international space station at the 10$^{-14}$ level \cite{Tino2013}. The SOC and I-SOC consortia studied the possibility to operate a space optical clock~\cite{Bongs2015,Origlia2016,Schiller2017}. Recently, the “Cold-Atom-Interferometry for future satellite gravimetry” (CAI) study~\cite{Trimeche2019} developed the necessary concepts of satellite gravimetry for geodesy purposes. Similar concepts were studied within mission proposals for space gravitational wave observatories~\cite{Loriani2019,Tino2019}.
In the frame of the ESA Cosmic Vision program (M3), the satellite test of the equivalence principle STE-QUEST was down-selected for a phase A study~\cite{Aguilera2014}. The outcome of this study validated the main concepts for such an operation with a dual-condensed source of Rb isotopes testing the UFF at the 2 10$^{-15}$ level. 
%The spacecraft orbit proposed back then was  highly elliptical in order to allow for a common view to ground stations connected by the MW link from the satellite contributing to a redshift test. Such a scenario made the integration time relatively long since important parts of the orbit were not practical for a the UFF measurement (weak gravity signal). 
The assessment of the mission was mainly critical towards TRL aspects of parts of the payload relative to the generation and manipulation of the ultra-cold ensembles. Since then, several of these limitations were overcome mainly thanks to the developments of national programs in France and Germany.

The ICE (Interf\'erometrie Coh\'erente pour l'Espace) project, funded by CNES is aiming to a WEP test with a dual species Rb/K atom interferometer on board of parabolic flights of 20\,s each. The experiment uses frequency-doubled telecom lasers to manipulate the atoms~\cite{Geiger2011}. Recently the payload was put on a 3-m microgravity simulator and produced an all-optical degenerate source of Rb at $35$\,nK tempreature allowing to explore the relevant weightlessness times of $400$\,ms~\cite{Condon2019}.

The DLR-funded consortium QUANTUS (QUANTen Gase Unter Schwerelosigkeit), aimed at developing transportable BEC sources capable of microgavity operation. The miniaturized devices were based on the atom chip technology on one hand and diode lasers on the other one. Different generations of these machines were operated at the 100-m high droptower at ZARM (Bremen). The first BEC under microgravity was demonstrated~\cite{vanZoest2010} as well as several key techniques relevant for an atomic UFF test such as seconds-long free expansions or long-time interferometers (675 ms)~\cite{Muntinga2013}. Moreover, an advanced version (QUANTUS-2 experiment) demonstrated a metrology-compatible duty cycle as short as $1.5$\,s for 10$^5$ BEC atoms of $^{87}$Rb~\cite{Rudolph2015}. The same payload was operating during the catapult mode at the Bremen droptower for $9$\,s achieving four complete cycles of BEC experiments in one shot. In view of a space interferometric test of the equivalence principle, the cold ensembles' expansion was slowed down using the delta-kick collimation technique to ultra-low energy levels of few tens pK~\cite{Rudolph2016}.  
This heritage made it possible in 2017 to create the first BEC in space on board of a sounding rocket built and operated by the MAIUS consortium~\cite{Becker2018}. During the short microgravity time of $6$\,min, more than 100 experiments central to matter-wave interferometry, including laser cooling and trapping of atoms in the presence of the large accelerations experienced during launch were executed. The results showed the maturity of miniaturized cold-atom technology for satellite-based implementation with respect to aspects such as reproducibility and autonomous operation.

In parallel, NASA developed the multi-user BEC facility Cold Atom Lab (CAL) aboard the International Space Station (ISS)~\cite{Elliott2018}. It provides a persistent orbital laboratory featuring as well a an atom chip-based system designed to create ultra-cold mixtures and degenerate samples of Rb and K. At the moment several consortia of researchers including US and German teams are conducting atom optics experiments with ${^{87}}$Rb condensates.

In the next five years, this fast pace bringing precision experiments featuring quantum gases will be kept. The second sounding rocket mission MAIUS-2 has the target to create quantum mixtures of Rb and K and to perform more experiments relevant to interferometry. Within the ICE project, the same pair of species will be operated in interferometric measurement campaigns on parabolic flights. A NASA-DLR joint mission will bring a successor of CAL (BECCAL) on board of the ISS and is currently under construction. Moreover, recent space-missions such as LISA-pathfinder \cite{Armano2018,Armano2019} and MICROSCOPE \cite{Touboul2017}, provide with a sound heritage on satellite control and drag-free operation aspects that are significant for an STE-QUEST-like mission. 

\subsection{Summary}

A dual species atom interferometer in space offers a new approach for testing the universality of free fall, complementary to classical tests.
The described scenario builds on heritage which demonstrated atom optics and atom interferometry on microgravity platforms as parabolic flights, a drop tower, and in space.
It anticipates a residual uncertainty in the \Eotvos ratio of $10^{-17}$ after 18\,months of integration.

\section{An advanced MICROSCOPE mission} \label{mscope}
\subsection{Heritage of the MICROSCOPE test of UFF/WEP} \label{sect:msc1}

MICROSCOPE \cite{touboul12} aimed to test UFF/WEP with an unprecedented uncertainty of $10^{-15}$. The T-SAGE (Twin Space Accelerometers for Gravitation Experiment) scientific payload, provided by ONERA, was integrated within a CNES micro-satellite. It was launched and injected into a 710\,km altitude, circular orbit, by a Soyouz launcher from Kourou on April 25, 2016. The orbit is sun-synchronous, dawn-dusk (i.e. the ascending node stays at 18\,h mean solar time) in order to have long eclipse-free periods (eclipses are defined as periods within the Earth{'}s shadow and happen only between May and July).
\subsubsection{Description of the experiment}
\label{sec:descr-exper}

T-SAGE \cite{touboul01} is composed of two parallel similar differential accelerometer instruments, each one with two concentric hollow cylindrical test-masses. They are exactly the same, except for the use of different materials for the test-masses. In one instrument (SUREF) the two test-masses have the same composition, and are made from a Platinum/Rhodium alloy (90/10). In the other instrument (SUEP) the test-masses have different compositions: Pt/Rh~(90/10) for the inner test-mass and Titanium/Aluminium/Vanadium~(90/6/4) (TA6V) for the outer test-mass (see Table~\ref{tab:masses}). 
\begin{table}[b]
\caption{\label{tab:masses}%
Main test-mass physical properties measured in the laboratory before integration in the instrument.
}
\begin{ruledtabular}
\begin{tabular}{lcccc}
\textrm{Measured}&
\textrm{SUREF}&
\textrm{SUREF}&
\textrm{SUEP} & SUEP\\
parameters & Inner mass & Outer mass & Inner mass & Outer mass \\
at $20\,^o$C & Pt/Rh & Pt/Rh & Pt/Rh & Ti/Al \\
\hline

Mass in kg & 0.401533 & 1.359813 & 0.401706 & 0.300939\\ \hline
Density in g\,cm$^{-3}$ & 19.967 & 19.980 & 19.972 & 4.420\\
%g\,cm$^{-3}$ &  &  &  & \\
\end{tabular}
\end{ruledtabular}
\end{table}

In the spirit, the experiment aims to compare the free fall of several test-masses orbiting the Earth. But, for practical reasons, the implementation is slightly more sophisticated and rests on two nested control loops.

The first loop is inside the payload T-SAGE constituted by 4 test-masses grouped by pairs in two differential accelerometers. Each test mass is placed between pairs of electrodes and its motion with respect to its cage fixed to the satellite is monitored by capacitive sensors. Then, this motion can be controlled  at rest by applying the appropriate electrostatic force calculated by a PID. This means that this electrostatic force compensates all other forces. In that way, the knowledge of the  applied  electrostatic potential allows to measure the acceleration $\overrightarrow\Gamma_i$ which would  affect the test-mass with respect to the satellite in absence of the  electrostatic force.
% A part of these accelerations are common to all test-masses and would be perfectly canceled in the differences of the measured acceleration in case of perfect instruments: this is the case, for example, of non-gravitational accelerations acting on the satellite but not on the test-masses.

 Noting $\overrightarrow\Gamma_i^{\rm app}$ the theoretical (modelled) acceleration applied to the mass $i$ and  $\overrightarrow\Gamma_i^{\rm meas}$ the corresponding measurement by the not perfect instrument, they can be linked by the simplified relation $\overrightarrow\Gamma_i^{\rm meas}= \overrightarrow K_{0,i} +  \left[M_i\right]\overrightarrow\Gamma_i^{\rm app}+ \overrightarrow \Gamma_i^{n}$ where $\overrightarrow K_{0,i} $ is a bias, the matrix $ \left[M_i\right]$ takes into account the scale factors and the alignment of the test-mass and $\overrightarrow \Gamma_i^{n}$ is the measurement noise. Introducing the common mode acceleration $\displaystyle \overrightarrow\Gamma_c^{\rm app}=\frac12\left(\overrightarrow\Gamma_1^{\rm app}+\overrightarrow\Gamma_2^{\rm app}\right)$ and the differential mode acceleration  $\displaystyle \overrightarrow\Gamma_d^{\rm app}=\frac12\left(\overrightarrow\Gamma_1^{\rm app}-\overrightarrow\Gamma_2^{\rm app}\right)$ with equivalent notations for $\overrightarrow\Gamma^{\rm meas}$, $\overrightarrow K_{0} $, $ [M]$ and $\overrightarrow \Gamma^{n}$, we get
     \begin{equation}
       \label{eq:1}
       \overrightarrow\Gamma_d^{\rm meas}= \overrightarrow K_{0,d} + \left[M_c\right]\overrightarrow\Gamma_d^{\rm app} +  \left[M_d\right]\overrightarrow\Gamma_c^{\rm app}+\overrightarrow \Gamma_d^{n}
     \end{equation}
With little algebra we can get an equivalent relation involving $\overrightarrow\Gamma_c^{\rm meas}$ instead of $\overrightarrow\Gamma_c^{\rm app}$:
\begin{equation}
  \label{eq:2}
  \overrightarrow\Gamma_d^{\rm meas}= \overrightarrow K'_{0,d} + \left[M'_c\right]\overrightarrow\Gamma_d^{\rm app} +  \left[M'_d\right]\overrightarrow{\Gamma'}_c^{\rm meas}+\overrightarrow {\Gamma'}_d^{n}
\end{equation}
Thus, $ \overrightarrow\Gamma_d^{\rm meas}$ represents an acceptable measurement of $\overrightarrow\Gamma_d^{\rm app} $ if:
\begin{itemize}
\item the bias is very low or well calibrated or  the knowledge of  the constant part (and very low frequency part) of $\overrightarrow\Gamma_d^{\rm app} $ is not mandatory;
\item the measurement noise is low enough;
\item the terms of the $ [M'_d]$ matrix are small and/or well calibrated and the common mode acceleration $\overrightarrow\Gamma_c^{\rm meas}$ is small and/or well-known;
\item the matrix$ [M'_c]$ is close to identity.
\end{itemize}
The other major control loop in the MICROSCOPE experiment is included in the Acceleration and Attitude Control System (AACS) which applies accelerations on the satellite in order to cancel (or at least to considerably reduce), the level of the common mode measured acceleration  $\overrightarrow\Gamma_c^{\rm meas}$. This is achieved by means of very performant cold gas thrusters. This system also ensures a very accurate control of the pointing as well as the angular velocity and acceleration based on the measurements of angular position by the stellar sensors and of the angular acceleration by T-SAGE.

It can be shown \cite{hardy13} than the differential mode  acceleration applied when the test-masses are controlled fixed with respect to the satellite can be written as
\begin{equation}
  \label{eq:3}
  \overrightarrow\Gamma_d^{\rm app}=\delta\left(2,1\right) \overrightarrow g \left( O_{sat} \right) +  \left( \left[ T \right]- \left[\rm{In}\right]  \right)\overrightarrow\Delta ,
\end{equation}
where
\begin{itemize}
\item $ \overrightarrow g \left( O_{sat} \right)$ is the gravity acceleration at the satellite level and $ \delta\left(2,1\right) =\frac{m_{g2}}{m_{i2}} - \frac{m_{g1}}{m_{i1}}$ is a good approximation of the E\"otv\"os parameter; thus $\delta\left(2,1\right) \overrightarrow g \left( O_{sat} \right)$ is the possible EP violation signal we are looking for;
\item  $\left[ T \right]$ is the gravity gradient tensor and  $\left[\rm{In}\right] $ is the matrix gradient of inertia which induce an acceleration proportional to the vector $\overrightarrow\Delta$ between the centres of the 2 test-masses.
\end{itemize}
Even if T-SAGE measures the linear acceleration along the 3 axes, the measurement along the X-axis which is also the axis of the cylindrical test-masses is the most accurate. Thus, in practice we mainly use the above equation projected on the X-axis. This axis is controlled, thanks to the AACS, parallel to the orbital plane and rotates with a frequency  $f_{\rm spin}$ around the Y-axis orthogonal to the orbital plane.   In these conditions the component $g_X$  of the gravity, and then the searched EP signal $\delta\left(2,1\right) g_X$, varies with a very stable frequency $f_{\rm EP} = f_{\rm orb} + f_{\rm spin}$ where $f_{\rm orb}$ is the mean orbital frequency of the satellite. The components $T_{XX}$ and  $T_{XZ}$ of the gravity gradient have magnitudes of about $1.5\, 10^{-6}$ s$^{-2}$ and are associated to components $\Delta_X$ and $\Delta_Z$ of the off-centerings which can be hardly smaller than 10 $\mu$m; this leads to a differential acceleration from gravity gradients of the order of  $10^{-11}$ ms$^{-2}$, much larger than the accuracy of $8\times 10^{-15}$ ms$^{-2}$ targeted for the EP signal. Hopefully, (i) this gravity gradient signal is mainly concentrated at DC and 2 $f_{\rm EP}$ frequencies, well decorralated from the EP signal and, (ii) the components $\Delta_X$ and $\Delta_Z$ can be accurately estimated in flight and the effect of the gradient can be corrected.

\subsubsection{First results}
\label{sec:first-results}

The MICROSCOPE measurements were organised in successive sessions having different goals. The longest sessions (up to 120 orbital periods, i.e. more than 700 000 s) were dedicated to the EP test. Other shorter sessions (typically 5 orbits) aimed to calibrate or control some characteristics of the experiment. Most of the time the two sensors, SUEP and SUREF, operated separately: the AACS controlled the common mode of the active sensor while the other sensor, undergoing larger accelerations due to the gravity gradient, was off.

The results of the analysis of the first EP sessions have been published end of 2017 \cite{Touboul2017}. One session on SUREF over 62 useful orbits allowed to check that, for this comparison of the free fall of two identical materials, no unexpected signal was present at the EP frequency; the result for the E\"otv\"os parameter was
\begin {equation}
\delta({\rm Pt,Pt})=[+4\pm{}4{\rm (stat)}] \times{}10^{-15} \quad (1\sigma \,\,  {\rm statistical \,\, uncertainty}). 
\end {equation}
Another session on SUEP over 120 orbits led to
\begin{equation}
\delta({\rm Ti,Pt})=[-1\pm{}9{\rm (stat)}\pm{}9{\rm (syst)}] \times{}10^{-15} \quad (1\sigma \,\,  {\rm statistical \,\, uncertainty}).
\end{equation}
This was already an improvement of one order of magnitude with respect to the best result obtained with the E\"otwash experiment \cite{Schlamminger2008}. In particular, there was no detection of any EP violation for titanium and platinum at this level of precision. It was also checked that the AACS and the metrology of the instrument behaved as expected. For example the estimation of the components of the off-centring between the 2 test-masses of the SUEP were:
\begin{equation}
  \label{eq:4}
  \Delta x = 20.14 \pm{}0.05 \,\mu{\rm m}, \qquad \Delta y = -7.4 \pm{}0.2 \,\mu{\rm m}, \qquad  \Delta z = -5.55 \pm{}0.05 \,\mu{\rm m}.
\end{equation}
The MICROSCOPE in orbit mission came to its end in October 2018. Additional scientific data are under validation and should improve the above result regarding both the statistical error and the level of systematics.

\subsection{Advanced Test of UFF/WEP in space} \label{sect:msc2}
Taking advantage of the MICROSCOPE mission heritage in paragraph \ref{sect:msc1}, a first return of experience has been performed in order to list the main error sources or limitation of performance. From that list, an improved concept is being developed. The objective of an advanced MICROSCOPE mission is to achieved a performance of $10^{-17}$ on the test of UFF/WEP, that's to say 100 times better than the former MICROSCOPE objective \cite{Touboul2017}. If the orbit is selected to the one of MICROSCOPE, this performance turns into a requirement of the accelerometric measurement resolution to $8\times{}10^{-17}$m\,s$^{-2}$ at EP measurement frequency, $f_{EP}$. If we consider (see below) an increased integration time of 480 orbits, the level of stochastic noise should be limited to about $10^{-13}$m\,s$^{-2}$Hz$^{-1/2}$ when discarding systematics, and 5 times better if the specification has to be split between systematic and stochastic errors with some margins. This specifications could be relaxed if several sessions of 480 orbits are performed. But we consider here to obtain the performance only on one session.

Others drivers to the mission definition are considered in order to enable more science than the Equivalence Principle test alone.

\subsubsection {Drivers from MICROSCOPE experience} \label{sect:msc2a}

The accelerometers of MICROSCOPE are processed by pair in order to extract the eventual EP violation in the difference of acceleration of two test-masses (section \ref{sect:msc1}). In the paper \cite{Touboul2017}, the measurement of the acceleration difference shows a noise of $0.5 \times{}10^{-10}$~m\,s$^{-2}$~Hz$^{-1/2}$ at $f_{EP}$=$3\times{}10^{-3}$~Hz. The other pair of test-masses (PtRh/PtRh) was better by a factor 5. This means that the instrument noise limited the performance of the experiment to the EP parameter to $2\times 10^{-15}$ over an integration time of about 1800 orbits (the total number of science orbits during the mission) for the pair of test-masses (TiAl/PtRh).

\begin{figure}
\center
\includegraphics[width=0.8\textwidth]{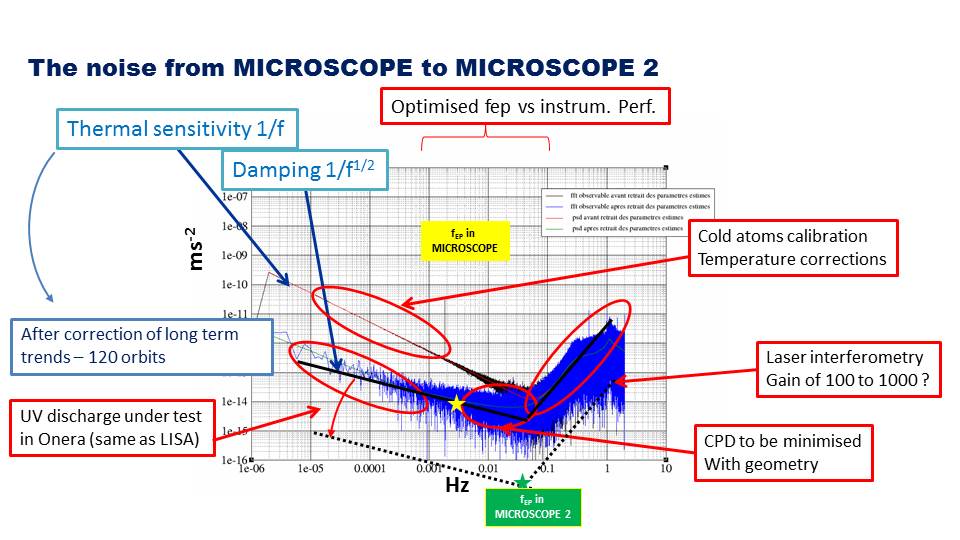}
\caption{Noise of the MICROSCOPE instrument with keys to improvement}
\label{fig_msc1}      
\end{figure}

The source of stochastic noise was mainly attributed to the gold wire which connects electrically the test-mass to a very stable reference voltage (Fig. \ref{fig_msc1}). The gold wire has a $7\rm\mu$m diameter and $\sim$ 25mm length. The mechanical stiffness, $k$, of the wire induces also a damping that is seen as a noise with frequency dependency $f^{-1/2}$. This noise varies like $k^{1/2}$ where the stiffness $k$ is proportional to $D^4/L^3$ for a wire of diameter D and length L. We consider here a stiffness due only to the mechanical flexion of the wire (that implies some constraints to the integration process). To reduce the noise by a factor 1000, the length should be increased by a factor 100 or the diameter reduced by a factor 33. In both cases, it appears far from what is technologically feasible today. A discharging device like the one foreseen for LISA \cite{Armano2018} is under study and may suppress this limitation.\\

At this lower level of resolution, others sources of stochastic noise, not dominant on MICROSCOPE, could appear: 
\begin{itemize}
\item Differential contact potentials \cite{speack99} have to be considered in an advanced MICROSCOPE mission. In MICROSCOPE the corresponding error budget gives an evaluation of $2\times{}10^{-13}$m\,s$^{-2}$Hz$^{-1/2}$. To reduce the effect of the contact potential over 480 orbits by a needed factor of 10, the gaps between the test-mass and the environment should be increased by a factor $10^{-1/3}\simeq 2$ as the effect is inversely proportional to the power 3 of the distance. This implies a reduction of control range by a factor 4 and an increase of the free-motion of the test-mass by a factor 2 as the stops limiting the free-motion should be taken away from the test-masses. This constraint can be handled.
\item The cracking of the Multi Layer Insulator of the satellite generates acceleration peaks with frequency occurrences higher than 0.01 Hz. This is a common mode signal that is removed in the case  of MICROSCOPE with the matching of the scale factors. The effect of the aliasing of this effect was evaluated to be negligible in the measurement bandwidth. However in an advanced version of the mission, one would like to increase the frequency of measurement and this effect should be considered properly in the design of the satellite. A stiffer structure may be a solution like the one selected for the ESA GOCE mission \cite{marque10}.
\end{itemize}

In addition to the stochastic noise, there is also the systematic errors. MICROSCOPE shows that by increasing the rotation of the satellite (to get a higher measurement frequency), the temperature stability becomes much better.

The measurement read-out is based on digital electronics. In MICROSCOPE the core of the digital servo-loop was performed with a DSP which calculates in 40bits accuracy and stores data in 32bits. It was shown that a peculiar attention has to be paid in the digital channel. If two orders of magnitude better is required that means at least an additional 7bits at all levels.
Other systematic errors may be improved like the star sensor noise or the drag-free control but with one order of magnitude as the resulting performance is well achieved in MICROSCOPE.
The operational process has also to be improved. In MICROSCOPE, the science sessions are limited to 120 orbits because the flight parameters of the satellite have to be adjusted every 15 days. The operation of an advanced mission should rely on longer sessions of e.g. 480 orbits instead of 120 to gain a direct factor 2 by the integration time only. An other operational limitation comes from the Moon that is dazzling the star sensor border once a month. Thus, the satellite has to be depointed in order to keep safe the SST control. This depointing induces thermal condition changes not compatible with the science requirements. Additional star sensors on the satellite in order to prevent a depointing of the satellite should be considered.

\subsubsection {New concept for an advanced MICROSCOPE mission} \label{sect:msc2b}

\begin{figure}
\center
\includegraphics[width=0.8\textwidth]{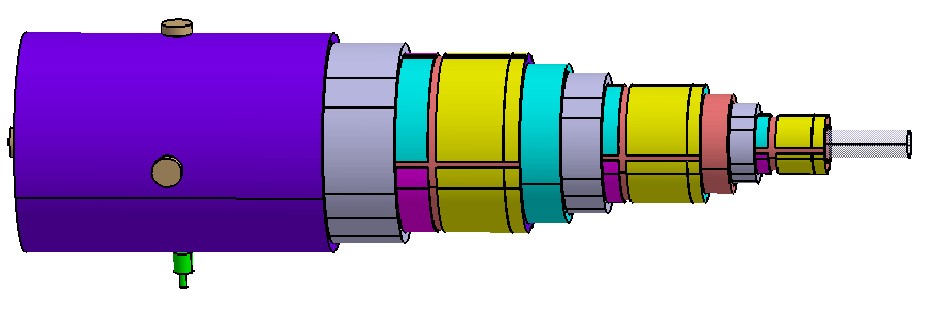}
\caption{Drawing of the 3 concentric test-masses (in grey) surrounded by the electrode of control (coloured parts).}
\label{fig_msc2}      
\end{figure}

In order to improve the performance, some parameters of the instrument have to be changed as listed in the previous paragraph. The core of the instrument could comprise 3 concentric test-masses as depicted on Fig. \ref{fig_msc2}. The choice of the materials has to be established in relation to theoretical considerations \cite{fayet19, Damour1994}. The choice of 3 different materials or of 2 identical materials within the 3 test-masses has to be evaluated. The advantage of 3 concentric test-masses is the ability to perform the drag-free control on the 3 test-masses at the same time and thus to realise two EP test with 2 pairs of material simultaneously. The comparison of the two results can be useful to improve the accuracy of the test and the rejection of common systematic errors. That was not possible on MICROSCOPE as the 2 pairs of material could not work simultaneously with the drag-free on both.

The principle of operation remains the same at the beginning of the test-mass control. A capacitive detector measures the test-mass motion. Then the information is taken into account by a digital servo loop controller that calculates the voltages necessary to apply to the electrodes in order to maintain the test-masses motionless. The applied voltages are representative of the acceleration measurement (see paragraph \ref{sect:msc1}). This is the basic principle of the electrostatic accelerometer.

Achieving a $10^{-17}$ means also improving the calibration process. In MICROSCOPE, the calibration was performed in inertial pointing which was not the nominal mode for the EP test. It was verified that the calibrated parameters allowed to reduced the common mode in all rotating configuration of the satellite. In an advanced mission, the calibration should be performed in inertial pointing and in rotating mode. The effect of the gravity gradient induced by the test-mass miss-centring should be negligible with a rotating spacecraft: in this case the EP signal and the gravity gradient are well decorrelated. The effect of the attitude motion has to be considered. With the electrostatic accelerometer it is possible to correct the in flight miss-centring after calibration by biasing the capacitive sensor. But the operation of the capacitive sensor out of the zero position induces larger sensitivity to temperature. To cope with this effect, it is proposed to use an interferometer position sensing in the servo-loop in science mode as performed in LISA \cite{armano16} with the test-mass cantered. This has several advantages:
\begin{itemize}
\item improvement of the acceleration noise at frequency higher than 0.01Hz,
\item suppression of the coupling in the loop between the electrostatic command and detection,
\item possibility to cancel the inertial motion effects (relaxation of miss-centring and attitude motion requirements),
\item possibility to perform a direct optical differential measurement between two test-masses.
\end{itemize}

As already mentioned, the test-mass charge has to be controlled by a discharging device. The example of LISA-Pathfinder or LISA gives a driver to the definition. However, in ONERA another option is under study and may enable a continuous discharge system.

To improve the measurement read-out, a new architecture of the digital controller has to be designed as the MICROSCOPE one is not compatible:
\begin{itemize}
\item The calculation has to be performed over 64 bits;
\item The operation frequency loop of 1027Hz in MICROSCOPE has to be re-considered with respect to the sampling frequency and the servo-loop bandwidth;
\item The operation frequency loop should be selected as a multiple of the sampling frequency and as a multiple of the satellite clock. This could minimise the risk of loosing performance in the averaging process of the 1027Hz data.
\end{itemize}

Extrapolating the MICROSCOPE figures, the sensor with the optical readout may have a volume of 50L, a mass of 50kg and a power consumption of 50W.

In order to extend the experimentation, and improve the performance in the low frequency range, an hybridisation of the electrostatic accelerometer with an atomic interferometer could be envisaged. The objective is to eliminate the long term drift that was estimated by a polynomial in the case of MICROSCOPE but that could be improved by an in-situ accelerometric measurement with cold atoms. This technology is under study and tested within the frame of an ESA gradiometer. In spite of the increased complexity, cold atoms may bring absolute scale factor determination of the electrostatic accelerometer. The optical readout can share the laser source with the atomic interferometer. The budget is not established but may be larger than the simple configuration instrument.

\subsubsection {Advanced MICROSCOPE additional science to be tested} \label{sect:msc2c}

The science to be tested is mainly the test of the Equivalence Principle.
The test of the Lorentz symmetry could be notably improved as it was already done with MICROSCOPE \cite{pihan17}. The study of a Chameleon fifth force is under evaluation in MICROSCOPE but should lead into recommendation to improve the sensitivity \cite{borras19} that could be considered in the future mission.
The addition of an optical position sensor could help in testing the Casimir effect by taking advantage of the capability of the electrostatic device to move the test mass accurately with a particular pattern \cite{lambrecht06}.
The use of cold atoms can extend the possible science to be tested:
\begin{itemize}
\item The comparison between macroscopic mass and quantum mass for an EP test;
\item Some tests of `big G' measurement with atoms are described in \cite{rosi16}. In the advanced mission, the 3 macroscopic masses can delivered a gravitational signal at a particular frequency that could improve the noise signal ratio;
\item Gradiometry experiment can also be undertaken with the cold atoms helped by the macroscopic test-mass calibrated motion. In this last case, the test-masses can be used either to cancel local gravity gradients or to generate a calibrated gravity gradient.
\end{itemize}


\newpage
\begin{thebibliography}{999}

\bibitem{Touboul2017} Touboul  \textit{et al.}, Phys. Rev. Lett. \textbf{119}, 231101 (2017)

\bibitem{Sumner2007} Sumner T. J., {\it et al.}, Advs. Space Res., \textbf{39}, 254-258 (2007).

\bibitem{Nobili2012}  Nobili A. M,. {\it et al.}, Class. Quant. Grav. \textbf{29}, 184011 (2012).

\bibitem{Reasenberg2014} Reasenberg, R. D., Class. Quant. Grav. \textbf{31}, 175013 (2014).

\bibitem{Amelino2009} Amelino-Camelia G. {\it et al.}, Experimental  Astronomy, \textbf{23}, 549-572, (2009).

\bibitem{Aguilera2014} 
Aguilera D. \emph{et al.}, Cassical and Quantum Gravity \textbf{{31}}, 115010 (2014).

\bibitem{Altschul2015} Altschul, B. \textit{et al.}, Advances in Space Research \textbf{55}, 501 - 524, (2015).

\bibitem{Armano2018} Armano, M. \textit{et al.}, Physical Review Letters \textbf{120}, 061101, (2018).

\bibitem{Armano2019} Armano, M. \textit{et al.}, Phys. Rev. D \textbf{99}, 082001, (2019).

\bibitem{Muntinga2013}
{M{\"u}ntinga} H. \emph{et al.}, Phys. Rev. Lett. \textbf{110}, 093602 (2013).

\bibitem{Becker2018}
Becker D. \emph{et al.}, Nature \textbf{562}, 391 (2018).

\bibitem{Geiger2011}
Geiger R., M{\'e}noret V., Stern G., Zahzam N., Cheinet P., Battelier B.,
  Villing A., Moron F., Lours M., Bidel Y., Bresson A., Landragin A. and Bouyer
  P., Nat. Comm. \textbf{2}, 474 (2011).

\bibitem{Roura2017}
Roura A., Phys. Rev. Lett. \textbf{118}, 160401 (2017).

\bibitem{Overstreet2017}
Overstreet C., Asenbaum P., Kovachy T., Notermans R., Hogan J.M. and Kasevich M.A., Phys. Rev. Lett. \textbf{120}, 183604 (2018)

\bibitem{Damico2017}
D'Amico G., Rosi G., Zhan S., Cacciapuoti L., Fattori M. and Tino G., Phys. Rev. Lett. \textbf{119}, 253201 (2017).

\bibitem{Cacciapuoti2009} Cacciapuoti, L. and Salomon, C., Eur. Phys. J. Special Topics, \textbf{172}, 57-68, (2009).

\bibitem{Ade2013}
Ade P. \textit{et al.}, The Planck Collaboration, arXiv: 1303.5076 (2013).

\bibitem{Schiff1960} Schiff L., Am.  J. Phys. \textbf{28}, 340 (1960).

\bibitem{Dicke1964} 
Dicke R.H., The Theoretical Significance of Experimental Relativity, Gordon and Breach, New York, 1964.

\bibitem{Thorne1973} Thorne K.S., Lee D.L. and Lightman A.P.,
  Phys. Rev. D \textbf{7}, 3563 (1973).

\bibitem{Will1993} Will C.M., \textit{Theory and experiment in
  gravitational physics, 2nd edition}, Cambridge U. Press (2018).

\bibitem{Jordan1946} Jordan P., Nature \textbf{164}, 637 (1946).

\bibitem{Brans1961} Brans C. and Dicke R., Phys. Rev. \textbf{124},
  925 (1961).

\bibitem{Damour2008} Damour T., in C. Amsler \textit{et al.} (Particle
  Data Group), Phys. Lett. B \textbf{667}, 1 (2008), 2010 update.

\bibitem{Hees2018} Hees, A.; Minazzoli, O.; Savalle, E.; Stadnik, Y. V. and Wolf, P. Phys. Rev. D \textbf{98}, 064051 (2018)

\bibitem{Taylor1988} Taylor T.R. and Veneziano G., Phys. Lett. B \textbf{213}, 450 (1988).

\bibitem{Damour1994} Damour T. and Polyakov A.M., Nucl. Phys. B
  \textbf{423}, 532 (1994).

\bibitem{Dimopoulos1996}
Dimopoulos S. and Giudice G., Phys. Lett. B \textbf{379}, 105 (1996).

\bibitem{Antoniadis1998} Antoniadis I., Dimopoulos S. and Dvali G.,
  Nucl. Phys. B \textbf{516}, 70 (1998).

\bibitem{Rubakov2001}
Rubakov V.A., Phys. Usp. \textbf{44}, 871 (2001).

\bibitem{Maartens2010}
Maartens R. and Koyama K., Living Rev. Relativity \textbf{13}, 5 (2010).

\bibitem{Adelberger2009} Adelberger E.G., Heckel B.R. and Nelson A.E.,
  Ann. Rev. Nucl. Part. Sci. \textbf{53}, 77 (2009).
  
\bibitem{Antoniadis2011} Antoniadis I., Baessler S., B\"uchner M.,
  Fedorov V.V., Hoedl S., Lambrecht A., Nesvizhevsky V.V., Pignol G.,
  Protasov K.V., Reynaud S. and Sobolev Yu., C. R. Physique
  \textbf{12}, 755 (2011).

\bibitem{Khoury2004a} Khoury J. and Weltman A.,
  Phys. Rev. Lett. \textbf{93}, 171104 (2004).

\bibitem{Khoury2004b} Khoury J. and Weltman A., Phys. Rev. D
  \textbf{69}, 044026 (2004).
  
\bibitem{Fayet2017} Fayet P., Phys. Rev. D \textbf{97}, 055039, (2018).

\bibitem{fayet19} Fayet P., Phys. Rev. D \textbf{99}, 055043, (2019).

\bibitem{Wetterich1988} Wetterich C., Nucl. Phys. B \textbf{302}, 668
  (1988).
  
\bibitem{Ratra1988}
Ratra B. and Peebles J., Phys. Rev. D \textbf{37}, 321 (1988).

\bibitem{Carroll1998} Carroll S. M., Phys. Rev. Lett. \textbf{81},
  3067 (1998).

\bibitem{Brax2004} Brax P., van de Bruck C., Davis A.-C., Khoury
  J. and Weltman A., Phys. Rev. D \textbf{70}, 123518 (2004).

\bibitem{Chiba2000} Chiba T., Okabe T. and Yamaguchi M., Phys. Rev. D
  \textbf{62}, 023511 (2000).
  
\bibitem{ArmendarizPicon2000} Armendariz-Picon C., Mukhanov V. and
  Steinhardt P.J., Phys. Rev. Lett. \textbf{85}, 4438 (2000).

\bibitem{ArmendarizPicon2001} Armendariz-Picon C., Mukhanov V. and
  Steinhardt P.J., Phys. Rev. D \textbf{63}, 103510 (2001).

\bibitem{Dvali2000} Dvali G.R., Gabadadze G. and Porrati M.,
  Phys. Lett. B \textbf{485}, 208 (2000).

\bibitem{Kamenshchik2001} Kamenshchik A.Y., Moschella U. and Pasquier
  V., Phys. Lett. B \textbf{511}, 265 (2001).

\bibitem{Bilic2002} Bilic N., Tupper G.B. and Viollier R.D.,
  Phys. Lett. B \textbf{535}, 17 (2002).
  
\bibitem{Bento2002} Bento M.C., Bertolami O. and Sen A.A.,
  Phys. Rev. D \textbf{66}, 043507 (2002).

\bibitem{Capozziello2011} Capozziello S. and De Laurentis M.,
  Phys. Reps. \textbf{509}, 167 (2011).

\bibitem{Nojiri2007} 
Nojiri S. and Odintsov S.D.,
  Int. J. Geom. Meth. Mod. Phys. \textbf{4}, 115 (2007). 

\bibitem{Caldwell2002} Caldwell R.R, Phys. Lett. B \textbf{545}, 23
  (2002).

\bibitem{Weinberg1989} Weinberg S., Rev. Mod. Phys. \textbf{61}, 1
  (1989).

\bibitem{Damour2012} Damour T., arXiv:1202.6311 (2012).

\bibitem{Muller2010} M\"uller H., Peters A. and Chu S., Nature
  \textbf{463}, 926 (2010).

\bibitem{Wolf2011} Wolf P., Blanchet L., Bord{\'e} Ch.J., Reynaud S.,
  Salomon C. and Cohen-Tannoudji C., Class. Quant. Grav. \textbf{28},
  145017 (2011).

\bibitem{Giulini2012} Giulini D., in Quantum Field Theory and Gravity,
  p. 345, Finster F. \textit{et al.} Eds., Springer (2012).

\bibitem{Schlamminger2008} Schlamminger S., Choi K.-Y., Wagner T.,
  Gundlach J. and Adelberger E., Phys. Rev. Lett. \textbf{100}, 041101
  (2008).

\bibitem{Tasson2011} Kostelecky, V. A. and Tasson, J. D., Phys. Rev. D, \textbf{83}, 016013, (2011).

\bibitem{Peters2001} Peters, A.; Chung, K. Y. and Chu, Metrologia \textbf{38}, 25, (2001).

\bibitem{Merlet2010} Merlet, S.; Bodart, Q.; Malossi, N.; Landragin, A.; Santos, F. P. D.; Gitlein, O. and Timmen, L., Metrologia \textbf{47}, L9-L11, (2010).

\bibitem{Schlippert2014} Schlippert D., Hartwig J., Albers H.,
  Richardson L.L., Schubert C., Roura A., Schleich W.P., Ertmer W.,
  and Rasel E.M., Phys. Rev. Lett. \textbf{112}, 203002 (2014).

\bibitem{Tarallo2014} Tarallo M.G., Mazzoni T., Poli N., Sutyrin D.V.,
  Zhang X., and Tino G.M., Phys. Rev. Lett. \textbf{113}, 023005, (2014).
  
\bibitem{Zhou2015} Zhou, L.; Long, S.; Tang, B.; Chen, X.; Gao, F.; Peng, W.; Duan, W.; Zhong, J.; Xiong, Z.; Wang, J.; Zhang, Y. and Zhan, M., Phys. Rev. Lett. \textbf{115}, 013004, (2015).

\bibitem{Overstreet2018} Overstreet, C.; Asenbaum, P.; Kovachy, T.; Notermans, R.; Hogan, J. M. and Kasevich, M. A., Physical Review Letters \textbf{120}, 183604, (2018).

\bibitem{Hartwig2015} Hartwig, J.; Abend, S.; Schubert, C.; Schlippert, D.; Ahlers, H.; Posso-Trujillo, K.; Gaaloul, N.; Ertmer, W. and Rasel, E. M., New Journal of Physics \textbf{17}, 035011, (2015).

\bibitem{Doser2010} Doser M., J. Phys.: Conf. Ser. \textbf{199},
  012009 (2010).
  
\bibitem{Perez2012} 
Perez P. and Sacquin Y., Classical and Quantum
  Gravity \textbf{29}, 184008 (2012).
  
\bibitem{Damour2010} Damour T. and Donoghue J.F., Phys. Rev. D
  \textbf{82}, 084033 (2010).
  
\bibitem{Arvanitaki2015} Arvanitaki, A.; Huang, J. and Van Tilburg, K., Phys. Rev. D \textbf{91}, 015015, (2015).

\bibitem{Stadnik2015} Stadnik, Y. V. and Flambaum, V. V., \textbf{115}, 201301, (2015).

\bibitem{pihan17} Pihan-Le Bars H. \textit{et al.}, 	arXiv:1705.11015 [gr-qc] (2017).

\bibitem{Pihan2019} Pihan-Le Bars H. \textit{et al.}, to be published.

\bibitem{YellowBook} Space-Time Explorer and QUantum Equivalence Space
  Test, Yellow Book of STE-QUEST, ESA/SRE \textbf{6} (2013).
  
\bibitem{Wolf2016} Wolf, P. and Blanchet, L., Classical and Quantum Gravity \textbf{33}, 035012, (2016).

\bibitem{Savalle2019} Savalle, E., Guerlin, C., Delva, P., Meynadier, F., Le Poncin-Lafitte, Cµ., and Wolf, P., Class. Quant. Grav., submitted, (2019), arXiv:1907.12320 .

\bibitem{Origlia2016} Origlia, S. {\it et al.}, Phys. Rev. A \textbf{98}, 053443 (2018).

\bibitem{Bongs2015}
Bongs K. \emph{et al.},  C. R. Phys. \textbf{{16}}, {553--564} ({2015}).

\bibitem{Schiller2017} Schiller, S. and Cacciapuoti, L., {I-SOC} {Scientific Requirements}, European Space Agency, document SCI-ESA-HRE-ESR-ISOC, http://www.exphy.uni-duesseldorf.de/PDF/SCI-ESA-HRE-ESR-ISOC\_Iss.1.1-Approved.pdf . 

\bibitem{Meynadier2018} Meynadier, F.; Delva, P.; le Poncin-Lafitte, C.; Guerlin, C. \& Wolf, P., Classical and Quantum Gravity \textbf{35}, 035018, (2018).

\bibitem{Panek2010} P. Panek, I. Prochazka and J. Kodet, Metrologia \textbf{47}, L13-L16 (2010). 

\bibitem{Prochazka2013} I. Prochazka, J. Kodet and J. Blazej, Rev. Sci. Instrum. \textbf{84}, 046107 (2013).

\bibitem{Tapley2004} Tapley, B. D., Bettadpur, S., Watkins, M., and Reigber, C., Geophys. Res. Lett., \textbf{31}, L09607, (2004).

\bibitem{Flury2008}
Flury J., Bettadpur S. and Tapley B.D., Adv. Space Res. \textbf{42}, 1414--1423 (2008).

\bibitem{Bock2011} Bock, H., J\"aggi, A., Meyer, U. et al., J. Geod \textbf{85}, 807, (2011).

\bibitem{Abich2019} K. Abich \textit{et al.}, Physical Review Letters \textbf{123}, 031101, (2019). 

\bibitem{Panet2018} Panet, I. ; Bonvalot, S.; Narteau, C.; Remy, D. \& Lemoine, J.-M., Nature geoscience \textbf{11}, 367-373, (2018).

\bibitem{Lion2017} Lion, G.; Panet, I.; Wolf, P.; Guerlin, C.; Bize, S. \& Delva, P., Journal of Geodesy, \textbf{91}, 597, (2017).

\bibitem{Denker2017} Denker, H.; Timmen, L.; Voigt, C.; Weyers, S.; Peik, E.; Margolis, H. S.; Delva, P.; Wolf, P. \& Petit, G., Journal of Geodesy \textbf{92}, 487, (2017).

\bibitem{Mehlstaeubler2018} Mehlstäubler, T. E.; Grosche, G.; Lisdat, C.; Schmidt, P. O. \& Denker, H., Rep. Prog. Phys. \textbf{81}, 064401, (2018).

\bibitem{Biancale2016} Biancale R. {\it et al.}, http://www3.mpifr-bonn.mpg.de/div/meetings/vonft/pdf-files/talks/E-GRASP\_Eratosthenes\_Biancale

\bibitem{Savoie2018}
Savoie D., Altorio M., Fang B., Sidorenkov L.A., Geiger R. and Landragin A., Sci. Adv. \textbf{4}, eaau7948 (2018).

\bibitem{Freier2016}
Freier C., Hauth M., Schkolnik V., Leykauf B., Schilling M., Wziontek H., Scherneck H.G., M\"uller J. and Peters A., J. Phys.: Conf. Ser. \textbf{723}, 012050 (2016).

\bibitem{Rosi2017}
Rosi G., D'Amico G., Cacciapuoti L., Sorrentino F., Prevedelli M., Zych M.,
Brukner \c C. and Tino G.M., Nat. Commun \textbf{8}, 15529 (2017).

\bibitem{Parker2018}
Parker R.H., Yu C., Zhong W., Estey B. and M\"uller H., Science \textbf{360}, 191-195 (2018).

\bibitem{Hohensee2013}
Hohensee M.A., M\"uller H. and Wiringa R.B., Phys. Rev. Lett. \textbf{111}, 151102 (2013).

\bibitem{Barrett2016}
Barrett B., Antoni-Micollier L., Chichet L., Battelier B., L\'ev\`eque T.,
  Landragin A. and Bouyer P., Nat. Commun. \textbf{7}, 13786 (2016).
  
\bibitem{Williams2016}
Williams J., Chiow S.w., Yu N. and M\"uller H., New J. Phys. \textbf{18}, 025018 (2016). 

\bibitem{Schubert2013}
Schubert C. \emph{et al.}, arXiv:1312.5963 (2013).

\bibitem{Blaser2001}
Blaser J.P., Classical and Quantum Gravity \textbf{18}, 2509--2514 (2001).

\bibitem{Nobili2016}
Nobili A.M., Phys. Rev. A \textbf{93}, 023617 (2016).

\bibitem{Lan2012}
Lan S.Y., Kuan P.C., Estey B., Haslinger P. and M\"uller H., Phys. Rev. Lett. \textbf{108}, 090402 (2012).

\bibitem{Kovachy2015}
Kovachy T., Hogan J.M., Sugarbaker A., Dickerson S.M., Donnelly C.A., Overstreet C. and Kasevich M.A., Phys. Rev. Lett. \textbf{114}, 143004 (2015).

\bibitem{Rudolph2016}
Rudolph J., \emph{{Matter-Wave Optics with Bose-Einstein Condensates in
  Microgravity}}, {dissertation, Leibniz Universit\"at Hannover} (2016).

\bibitem{Cheinet2008}
Cheinet P., Canuel B., Pereira dos~Santos F., Gauguet A., Yver-Leduc F. and
  Landragin A., IEEE Trans. Instrum. Meas. \textbf{57}, 1141--1148 (2008).

\bibitem{Langlois2017}
Langlois M., Caldani R., Trimeche A., Merlet S. and Pereira~dos Santos F., Phys. Rev. A \textbf{96}, 053624 (2017).

\bibitem{Richardson2019}
Richardson L.L. \emph{et al.}, arXiv:1902.02867 (2019).

\bibitem{LeGouet2007}
Gou\"et J.L., Cheinet P., Kim J., Holleville D., Clairon A., Landragin A. and
  Pereira dos Santos F., The European Physical Journal D \textbf{44}, 419--425 (2007).

\bibitem{LouchetChauvet2011}
Louchet-Chauvet A., Farah T., Bodart Q., Clairon A., Landragin A., Merlet S.
  and Pereira dos Santos F., New J. Phys. \textbf{13}, 065025 (2011).

\bibitem{Debs2011}
Debs J.E., Altin P.A., Barter T.H., D\"oring D., Dennis G.R., McDonald G.,
  Anderson R.P., Close J.D. and Robins N.P., Phys. Rev. A \textbf{84}, 033610 (2011).

\bibitem{Haslinger2018}
Haslinger P., Jaffe M., Xu V., Schwartz O., Sonnleitner M., Ritsch-Marte M.,
  Ritsch H. and M\"uller H., Nat. Phys. \textbf{14}, 257 (2018).  
  
\bibitem{Nicholson2015}
Nicholson T.L. \emph{et al.}, Nat. Commun. \textbf{6}, 6896 (2015).  
  
\bibitem{Thalhammer2008}
Thalhammer G., Barontini G., De~Sarlo L., Catani J., Minardi F. and Inguscio M., Phys. Rev. Lett. \textbf{100}, 210402 (2008).

\bibitem{Ferrari2002}
Ferrari G. \emph{et al.}, Phys. Rev. Lett. \textbf{89}, 053202 (2002).

\bibitem{Leveque2009}
L{\'e}v{\`e}que T., Gauguet A., Michaud F., Pereira dos~Santos F. and Landragin
  A., Phys. Rev. Lett. \textbf{103}, 080405 (2009).

\bibitem{Hogan2008}
Hogan J.M., Johnson D.M.S. and Kasevich M.A., arXiv:0806.3261 (2008).

\bibitem{Bongs2006}
Bongs K., Launay R. and Kasevich M.A., Appl. Phys. B \textbf{84}, 599 (2006).

\bibitem{Borde2004}
Bord{\'e} C., Gen. Relativ. Gravit. \textbf{36}, 475--502 (2004).

\bibitem{Schuldt2015}
Schuldt T. \emph{et al.}, Experimental Astronomy \textbf{39}, 167--206 (2015).

\bibitem{Rudolph2015}
Rudolph J. \emph{et al.}, New J. Phys. \textbf{17}, 065001 (2015).

\bibitem{Jentsch2004} Jentsch C. \emph{et al.}, General Relativity and Gravitation \textbf{36}, 2197 (2005).

\bibitem{Sorrentino2010}
Sorrentino F. \emph{et al.}, {Microgravity Sci. Technol.} \textbf{{22}}, {551--561} ({2010}).

\bibitem{Tino2013} Tino G.~M., Nuclear Physics B (Proc. Suppl.) \textbf{243}, 203 (2013).

\bibitem{Trimeche2019}
Trimeche A. \emph{et al.}, arXiv:1903.09828 (2019).

\bibitem{Loriani2019}
Loriani S. \emph{et al.}, New J. Phys. \textbf{21}, 063030 (2019).

\bibitem{Tino2019}
M.~Tino G. \emph{et al.}, arXiv:1907.03867
 (2019).
 
\bibitem{Condon2019}
Condon G., Rabault M., Barrett B., Chichet L., Arguel R., Eneriz-Imaz H., Naik
  D., Bertoldi A., Battelier B., Bouyer P. and Landragin A., arXiv:1906.10063 (2019).

\bibitem{vanZoest2010}
{van Zoest} T. \emph{et al.}, Science \textbf{328}, 1540 (2010).

\bibitem{Elliott2018}
Elliott E.R., Krutzik M.C., Williams J.R., Thompson R.J. and Aveline D.C., npj Microgravity \textbf{4}, 16 (2018).

\bibitem{touboul12} Touboul P. \textit{et al.}, Class. and Quant. Grav. \textbf{29}, 184010 (2012).
\bibitem{touboul01} Touboul P. \textit{et al.}, Class. and Quant. Grav. \textbf{18}, 2487-2498 (2001).
\bibitem{hardy13} Hardy E. \textit{et al.}, Adv. Space Res. \textbf{52}, 1634 (2013).
%\bibitem{touboul17} Touboul P., et al, Phys. Rev. Lett. \textbf{109}, 231101 (2017).
%\bibitem{schlamminger08} Schlamminger S., et al., Phys. Rev. Lett. \textbf{100}, 041101 (2008).
%\bibitem{armano18} Armano M., et al, Phys. Rev. D \textbf{98}, 062001 (2018).
\bibitem{armano16} Armano M. \textit{et al.}, Phys. Rev. Lett. \textbf{116}, 231101 (2016).
\bibitem{speack99} Speak C., Class. and Quant. Grav. \textbf{13}, 11A/039 (1999).
\bibitem{marque10} Marque J.P. \textit{et al.}, ESA Living Planet Symposium, ESA Special Publication \textbf{686}, 57 (2010).
%\bibitem{damour94} Damour. T., et al, Nuclear Physics B \textbf{423}, 532-558 (1994).
\bibitem{borras19} Pernot-Borras M. \textit{et al.}, in preparation (2019).
\bibitem{lambrecht06} Lambrecht A. \textit{et al.}, New Jour. of Phys. \textbf{8}, 243 (2006).
\bibitem{rosi16} Rosi G., J. Phys B. Mol. Opt. Phys. \textbf{49}, 202002 (2016).
 


\end{thebibliography}
\end{document}